\documentclass[aps,prx,twocolumn,superscriptaddress,longbibliography]{revtex4-2}
\usepackage{graphicx}
\usepackage{bm}
\usepackage{natbib}
\usepackage{color}
\usepackage{epstopdf}
\usepackage{amsmath} 
\usepackage{url}
\definecolor{darkGreen}{RGB}{0,110,0}
\definecolor{darkBlue}{RGB}{0,0,130}
\usepackage[colorlinks=true, urlcolor=darkBlue,citecolor=darkGreen,linkcolor=darkBlue]{hyperref}

\begin{document}

\title{Topological squashed entanglement: nonlocal order parameter for one-dimensional topological superconductors}


\author{Alfonso Maiellaro}
\affiliation{Dipartimento di Ingegneria Industriale, Università di Salerno, Via Giovanni Paolo II, 132, I-84084 Fisciano (SA), Italy}
\author{Antonio Marino}
\affiliation{Dipartimento di Ingegneria Industriale, Università di Salerno, Via Giovanni Paolo II, 132, I-84084 Fisciano (SA), Italy}
\author{Fabrizio Illuminati}
\affiliation{Dipartimento di Ingegneria Industriale, Università di Salerno, Via Giovanni Paolo II, 132, I-84084 Fisciano (SA), Italy}
\affiliation{INFN, Sezione di Napoli, Gruppo collegato di Salerno,Italy}


\date{July 4, 2022}

\begin{abstract}
Identifying entanglement-based order parameters characterizing topological systems, in particular topological superconductors and topological insulators, has remained a major challenge for the physics of quantum matter in the last two decades. Here we show that the end-to-end, long-distance, bipartite squashed entanglement between the edges of a many-body system, defined in terms of the edge-to-edge quantum conditional mutual information, is the natural nonlocal order parameter for topological superconductors in one dimension as well as in quasi one-dimensional geometries. For the Kitaev chain in the entire topological phase, the edge squashed entanglement is quantized to $\log(2)/2$, half the maximal Bell-state entanglement, and vanishes in the trivial phase. Such topological squashed entanglement exhibits the correct scaling at the quantum phase transition, is stable in the presence of interactions, and is robust against disorder and local perturbations. Edge quantum conditional mutual information and edge squashed entanglement defined with respect to different multipartitions discriminate topological superconductors from symmetry breaking magnets, as shown by comparing the fermionic Kitaev chain and the spin-1/2 Ising model in transverse field. For systems featuring multiple topological phases with different numbers of edge modes, like the quasi 1D Kitaev ladder, topological squashed entanglement counts the number of Majorana excitations and distinguishes the different topological phases of the system. In fact, we show that the edge quantum conditional mutual information and the edge squashed entanglement remain valid detectors of topological superconductivity even for systems, like the Kitaev tie with long-range hopping, featuring geometrical frustration and a suppressed bulk-edge correspondence.
\end{abstract}


\maketitle


\section{Introduction}
\label{introduction} 

Condensed matter physics is witnessing, among others, two groundbreaking and concurring developments, respectively the application of concepts and methods of quantum information science and the investigation of topological phases of quantum matter. 

Via the identification of entanglement boundary laws and their violations, bipartite block entanglement (as measured by the reduced von Neumann entropy in a bipartite system) has become a central tool for the characterization and the diagnostics of large classes of phenomena in quantum many-body physics \cite{Wen2019,Laflorencie2016,Carr2010,Eisert2010,Amico2008}. 

At the same time, by featuring perfectly conducting edge modes, patterns of long-distance entanglement and robust ground-state degeneracy without symmetry breaking, three traits that generalize the concept of ordered phase and phase transition beyond the Ginzburg-Landau paradigm, topological states of matter have attracted increasing attention both for their fundamental interest and their potentiality for applications \cite{Wen2019,Laflorencie2016,Carr2010,Girvin2019,Ozawa2019,Chiu2016,Qi2011}. 

In particular, topological superconductors hosting edge Majorana zero energy modes (MZEMs) \cite{Sato2017} have been proposed as the working principle of various disruptive quantum technologies, including fault-tolerant topological quantum computation \cite{Pachos2012,Alicea2012,Beenakker2013}. 

Topological order in two-dimensional systems is identified and detected by the sub-leading contribution to the bipartite bulk-boundary von Neumann entanglement entropy, the so-called topological entanglement entropy (TEE)~\cite{Hamma2005,Kitaev2006,Levin2006}. This is a great success of entanglement theory applied to the investigation of quantum matter. 

Yet, the approach based on block entropies and simple bipartitions of a system into two connected parts (also often named "halves" or "blocks") is limited and cannot be applied to various important instances, thus showing the need for more general and advanced methods of entanglement theory. Indeed, in one dimension at zero temperature, no measure of bipartite entanglement based on simple bipartitions, including the entanglement spectrum, can discriminate topologically ordered phases from those corresponding to symmetry-breaking order \cite{Pollmann2010}. 

Moreover, irrespective of the spatial dimension, TEE cannot be generalized to finite-temperature and nonequilibrium processes since the von Neumann entropy, when defined on mixed states, includes contributions both from quantum and classical correlations, thus ceasing to be a genuine, {\it bona fide} measure of entanglement and nonclassicality.

A further limitation of the approach based on the elementary bipartition of the system in two connected parts (or halves, or blocks) is that it is not suitable for the characterization and quantification of the nonlocal correlations that are in place between arbitrary subsystems, either connected or disconnected; in particular, block entanglement entropy provides no information on the physics of edge correlations and cannot characterize and quantify the long-distance entanglement between edge modes. Therefore, going beyond the approach based on simple bipartitions by introducing bipartite entanglement measures defined on multipartitions would allow to identify richer conceptual structures and investigate broader classes of complex physical phenomena.

Thus motivated, in the present work we discuss a multipartition-based measure of bipartite entanglement, the squashed entanglement (SE), previously introduced in the general context of quantum information theory, and we apply it to the study of one-dimensional and quasi one-dimensional topological superconductors with model Hamiltonians supporting Majorana zero energy modes (MZEMs) at the system edges. We show that the long-distance squashed entanglement SE between the edges, that we name topological squashed entanglement (TSE), is the nonvanishing, \emph{nonlocal} order parameter that 1) characterizes unambiguously the topologically ordered phases of topological superconductors, 2) discriminates topological order from Ginzburg-Landau order associated to spontaneous symmetry breaking and nonvanishing local order parameters, and 3) distinguishes between different types of topologically ordered phases. In order to avoid confusion with the standard nomenclature about (local) order parameters, from now onward we will denote by topological order parameter or order parameter for topological superconductivity a nonlocal quantity that fulfils all the above criteria, being also very robust under even strong variations of the sample conditions (disorder or any local perturbation) and scaling exponentially with the system size to an asymptotic topological invariant. The latter is a numerical constant measurable with arbitrary precision and defined in terms of bulk Hamiltonians. The nonlocal and quantized nature of such a quantity is the main difference with order parameters classifying other phases of matter and standard Ginzburg-Landau order associated to spontaneous symmetry breaking. 

As we will show, one can define two basic forms of upper bounds on bipartite SE, respectively in terms of the tripartitions and quadripartitions that correspond to two different forms of the bipartite quantum conditional mutual information (QCMI). This general property of SE immediately allows to introduce two forms of the corresponding TSE order parameter, the first one based on edge-bulk-edge tripartitions, the second one based on edge-bulk-bulk-edge quadripartitions. We will then show how the latter identifies unambiguously the topological regime in the Kitaev fermionic chain, discriminates it from standard Ginzburg-Landau order in the spin-$1/2$ Ising chain, is stable in the presence of interactions, and is robust against the effects of disorder. Further, we will discuss how the TSE discriminates between different topological phases in fermionic systems defined on quasi one-dimensional geometries of higher complexity such as two--leg Kitaev ladders. Finally, we will consider models with long--range hopping and consequently suppressed bulk--boundary correspondence, such as the Kitaev tie, and we will show that even in this case the TSE is a good identifier of topological features.

In the present work we focus on the thorough investigation of edge QCMI and edge SE to three paradigmatic instances of $1D$ and quasi-$1D$ topological systems. In the conclusions, concerning future perspectives, we will discuss how to generalize the present framework to the study of the edge entanglement structure of many-body systems in higher dimensions. 

The paper is organized as follows. In Section \ref{squashed} we introduce the QCMI and the SE, review their main properties, and introduce the two fundamental forms of upper bounds on the SE based on the QCMI. In Section \ref{topoentang} we review the basic models of topological superconductivity in different lattice geometries and define the different forms of the TSE order parameter corresponding to tripartitions and quadripartitions. In Section \ref{KC} we discuss in detail the TSE of the $1D$ Kitaev model for topological superconductivity, compare it to the end-to-end SE in the Ising chain, and discuss the effects of interactions and disorder. In Section \ref{KL} we study the TSE of the two--leg Kitaev ladder and show how it identifies and distinguishes between the different topological phases of the model, while in Section \ref{KT} we carry out the same investigation for the case of the Kitaev tie with long--range hopping and suppressed bulk--boundary correspondence. Finally, in Section \ref{Concl} we review our results and consider possible generalizations and further applications of SE in the study of quantum matter.

\section{Squashed entanglement}
\label{squashed}

\subsection{Definition and fundamental properties}

We are looking for a measure of bipartite entanglement that has the following properties:

\begin{enumerate}

\item It is defined in any spatial dimension, at any temperature, and on all quantum states (pure or mixed).

\item It is bipartite but generically defined in terms of non-overlapping, distinguishable multipartitions, so that it can be defined on pairs of any two subsystems, either connected or disconnected, and must reduce to the bipartite von Neumann entanglement entropy on pure states of simple bipartitions.

\item It is a true, {\it bona fide} measure of bipartite entanglement, i.e. a convex entanglement monotone that in addition is also asymptotically continuous, monogamous, and additive on tensor products \cite{Horodecki2009,Bengtsson2020}.

\end{enumerate}

In fact, such a measure exists and is the so-called squashed entanglement (SE) \cite{Tucci2002,Christandl2004}. 
Given a quantum state $\rho_{AB}$ of a bipartite quantum system $AB$, the SE $E_{sq}(\rho_{AB})$ between subsystems $A$ and $B$ in 
state $\rho_{AB}$  is defined as:
\begin{equation}
E_{sq}(\rho_{AB}) = \inf_{\rho_{ABC}} \bigg\{ I(A:B|C) \bigg\} \, ,
\label{SE}
\end{equation}
where the infimum is taken over all the quantum-state extensions of arbitrary size $\rho_{ABC}$ such that $\rho_{AB}=Tr_{C}(\rho_{ABC})$, and $I(A:B|C)$ is the quantum conditional mutual information (QCMI) between subsystems $A$ and $B$ conditioned by extension $C$: 
\begin{equation}
I(A:B|C) = \frac{1}{2} \big[ S(\rho_{AC}) + S(\rho_{BC}) - S(\rho_{C}) - S(\rho_{ABC}) \big] ,
\label{QCMI}
\end{equation}
where $\rho_{AC} = Tr_{B}(\rho_{ABC})$, $\rho_{BC} = Tr_{A}(\rho_{ABC})$, $\rho_{C} = Tr_{AB}(\rho_{ABC})$, and $S(\rho)$ denotes the von Neumann entropy of a given quantum state $\rho$ \cite{Bengtsson2020}. Notice that here and in the following, for a matter of convenience, we adopt a definition of the QCMI that differs slightly, by a prefactor $1/2$, from the original one \cite{Cerf1997,Cerf1998}; consequently, the same factor is conveniently absorbed in our definition of the SE.

SE owes its name by the construction Eqs. \ref{SE}--\ref{QCMI} that "squashes" out the classical correlations to leave only the quantum contributions to the mutual information between parties $A$ and $B$ conditioned by party $C$ (that in informatic terms can be seen as the "conditioning environment"). SE is a lower bound on the entanglement of formation and an upper bound on the distillable secret key or distillable entanglement of a quantum state or a quantum channel \cite{Christandl2004,Wilde2016}, and reduces to the von Neumann entanglement entropy on pure states of a bipartite system $AB$. Indeed, if $\rho_{AB}$ is pure, then $\rho_{ABC} = \rho_{AB} \otimes \rho_{C}$, and thus $E_{sq}(\rho_{AB}) = [S(\rho_{A}) + S(\rho_{B})]/2 = S(\rho_{A}) = S(\rho_{B})$.

SE is unique in that it is the only known entanglement quantifier that enjoys all the desirable axiomatic properties required for a {\it bona fide} measure of entanglement \cite{Bengtsson2020}. 

Firstly, SE is a full entanglement monotone, i.e. non increasing under local operations and classical communication (LOCC), and convex~\cite{Christandl2004}: $E_{sq} (\lambda \rho + (1 - \lambda) \sigma) \leq \lambda E_{sq}(\rho) + (1 - \lambda) E_{sq} (\sigma)$, with $\lambda \in [0,1]$. 

Moreover, SE enjoys important additional properties that promote it to a full entanglement measure. To begin with, SE is additive on tensor products~\cite{Christandl2004}: $E_{sq}(\rho \otimes \sigma) = E_{sq}(\rho) + E_{sq}(\sigma)$. Additivity is a very important requirement: the total entanglement of a global state that is the tensor product of states of independent (uncorrelated) systems must be the sum of each individual entanglement. This property must hold for entanglement and entropy just like any other extensive physical quantity. Entanglement monotones that fail to be additive, like the entanglement negativity, can be at best only approximate quantifiers of quantum entanglement.

SE is also continuous~\cite{Alicki2004}: if two sequences of states $\rho_{m}$ and $\sigma_{m}$ converge in trace norm: $\lim_{m \rightarrow \infty} \Vert \rho_{m} - \sigma_{m} \Vert_{1} = 0$, then $\lim_{m \rightarrow \infty} E_{sq}(\rho_{m}) - E_{sq}(\sigma_{m}) = 0$. Continuity is also a very important requirement: two states that are infinitely close in Hilbert space according to a contractive norm must differ infinitesimally in their physical properties, including entanglement. Many entanglement quantifiers that are widely popular and used because of their straightforward computability, like the entanglement negativity, do not enjoy continuity.

Furthermore, SE is faithful, that is, $E_{sq} (\rho_{AB}) \geq 0$, and $E_{sq} (\rho_{AB}) = 0$ if and only if $\rho_{AB}$ is separable \cite{Brandao2011}. Again, this is a very important property because it guarantees that a vanishing SE implies with certainty that a quantum state is separable. Unfaithful entanglement quantifiers can be zero on entangled states, the paramount example being, once more, the entanglement negativity.

Finally, SE satisfies entanglement monogamy \cite{Coffman2000}; given three parties $A$, $B$, and $C$, the following monogamy inequality holds \cite{Koashi2004}: $E_{sq} (\rho_{ABC}) \geq E_{sq} (\rho_{AB}) + E_{sq} (\rho_{AC})$, that is, bipartite SE cannot be freely shared among multiple parties; in particular, if $A$ is maximally entangled with $B$, then it cannot be entangled with $C$. This property assures that one can introduce generalizations of SE in order to measure multipartite entanglement beyond the bipartite one. Moreover, the fact that SE is monogamous has important consequences on the structure of the bipartite SE between bulk and edge modes that we will study later on in this work (We note in passing that the above monogamy inequality can be extended straightforwardly to an arbitrary number $N$ of parties \cite{Osborne2006}).

SE is an entanglement measure with very important operational meaning. In quantum communication theory it is defined in terms of the communication cost in the distribution of quantum states among multiple parties \cite{Devetak2008} and it is a tight upper bound on the length of a secret key shared by two parties holding many copies of a quantum state \cite{LiWinter2014,Wilde2016}. Moreover, it allows for multipartite extensions with operational meaning \cite{Yang2009}. Finally, SE plays a fundamental role in channel theory, as the SE of a quantum channel is the optimal upper bound on the quantum communication capacity of any channel assisted by unlimited classical communication \cite{Takeoka2014}. In particular, SE provides the tightest known bound for this type of capacity (also called two-way assisted quantum capacity) for the case of the amplitude damping channel \cite{Pirandola2017}.

Because of all the above properties, SE has been dubbed the "perfect" measure of entanglement \cite{Bengtsson2020}, the only serious drawback concerning its computability. In fact, although computing SE has been shown to be an NP-complete problem \cite{Huang2014}, nevertheless it has been calculated analytically for some nontrivial classes of states \cite{Christandl2005,DePalma2019}. Moreover, and perhaps more importantly, SE enjoys a set of very useful lower bounds in terms of the reduced von Neumann entropies, the relative entropy of entanglement and the relative 2-R\'enyi entropy \cite{CarlenLieb2012,LiWinter2014,Brandao2015,FawziRenner2015}. These lower bounds can be combined with the natural upper bounds in terms of the tripartite and quadripartite quantum conditional mutual information (QCMI) that we will introduce in the following, in order to provide a tight quantitative characterization of SE.

\subsection{From bipartitions and two blocks to multipartitions and any pair of subsystems}

Given a physical system $G$ in a pure state $\rho_{G} = |G \rangle \langle G |$, for instance a ground state in condensed matter physics and quantum statistical mechanics or a vacuum state in quantum field theory, one typically focuses on simple bipartitions $G=AB$ of system $G$ in {\it two connected subsystems} (or blocks, or parts, or halves) $A$ and $B$, and considers then the pure--state bipartite entanglement between the two subsystems as quantified by the so-called von Neumann block entanglement entropy $S (\rho_{A}) = S (\rho_{B})$ of the reduced local states $\rho_{A} = Tr_{B} \rho_{G}$ or, equivalently, $\rho_{B} = Tr_{A} \rho_{G}$. 

Squashed entanglement greatly extends this picture to include all possible different forms of bipartite entanglement, including the case of mixed states and disconnected subsystems. To set the stage, consider a quantum system $G$ in an arbitrary state, pure or mixed, $\rho_{G}$. Next, consider partitioning the global system $G$ in any two subsystems $A$ and $B$ plus a reminder $C$: $G=ABC$. In turn, when suitable, one may consider further partitioning the remainder $C$ as well, as we will see in the following. When $C$ is the empty set and $\rho_{G}$ is a pure state, one recovers the standard two--block bipartition and SE reduces to the von Neumann block entanglement entropy.

We now wish to determine the SE $E_{sq}(\rho_{AB})$ existing between any pair of subsystems $A$ and $B$, possibly disconnected, in the reduced state $\rho_{AB} = Tr_{C} \rho_{G}$. The first important observation here in order is that there are always two, and only two, equivalent ways to obtain the same reduced state $\rho_{AB}$ from the global state $\rho_{G}$ that however give rise to two different (non equivalent) expressions for the QCMI $I$ of Eq.~\ref{QCMI} prior to the extremization procedure in Eq.~\ref{SE} that defines the (unique) SE $E_{sq}(\rho_{AB})$. These two different expressions of the QCMI define two fundamental upper bounds on the true SE that may in principle be different. The two expressions obviously give rise to the same unique SE once extremization is performed in Eq.~\ref{SE}. 

Indeed, besides the immediate {\emph{tripartite}} form $\rho_{G} = \rho_{ABC}$ such that $\rho_{AB} = Tr_{C}(\rho_{ABC})$, one can consider a further bipartite splitting of the reminder: $C = C_1C_2$ and obtain the corresponding {\emph{quadripartite}} form 
$\rho_{G} = \rho_{ABC} = \rho_{ABC_1C_2}$ such that $\rho_{AB} = Tr_{C_1}(\rho_{ABC_1})$ (or, equivalently, $\rho_{AB} = Tr_{C_2}(\rho_{ABC_2})$), where in turn $\rho_{ABC_1} = Tr_{C_2}(\rho_{ABC_1C_2})$ (or, equivalently, $\rho_{ABC_2} = Tr_{C_1}(\rho_{ABC_1C_2})$). The two different procedures to derive the reduced state $\rho_{AB}$ give rise in Eq.~\ref{QCMI} to two different QCMIs $I_{(3)}(A:B|C)$ and $I_{(4)}(A:B|C_1)$ (or, equivalently, $I_{(4)}(A:B|C_2)$) defined with respect to the tripartition $ABC$ and to the quadripartition $ABC_1C_2$, respectively. In explicit form, the two expressions read:
\begin{eqnarray}
	\label{upper3}
&& I_{(3)} = \frac{1}{2} \biggl[ S(\rho_{AC}) + S(\rho_{BC}) - S(\rho_{ABC}) - S(\rho_{C})\biggr] \, , \\
&& \nonumber \\
\label{upper4}
&& I_{(4)} = \frac{1}{2} \biggl[S(\rho_{AC_1}) + S(\rho_{BC_1}) - S(\rho_{ABC_1}) - S(\rho_{C_1})\biggr] ,
\end{eqnarray}
where $S(\rho_{ABC}) = 0$ if $\rho_{ABC} = \rho_{G}$ is pure. 

\begin{figure*}
\includegraphics[scale=0.14]{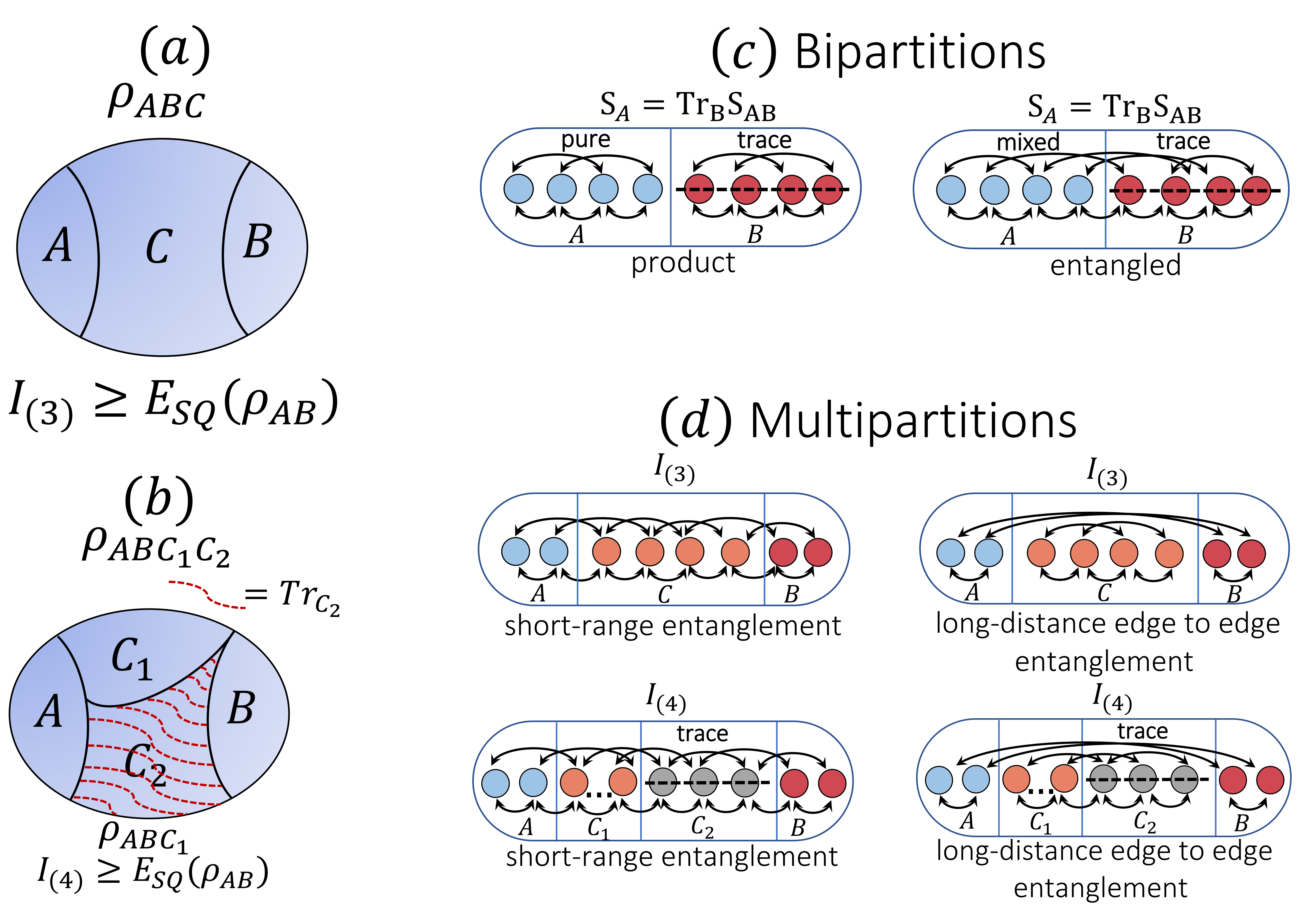}
\caption{Panels (a) and (b): Construction pattern of the two upper bounds, the QCMIs $I_{(3)}(A:B|C)$ and $I_{(4)}(A:B|C_1)$, on the true SE between two arbitrary subsystems $A$ and $B$ of a given quantum system partitioned, respectively as $ABC$ and as $ABC_1C_2$. In the given example the two subsystems $A$ and $B$ are disconnected, but the construction holds in general for any pair of non-overlapping subsystems, either connected or disconnected. Panel (c): A system in a globally pure state is partitioned in two connected subsystems (blocks); after tracing over the degrees of freedom of one of the blocks, the von Neumann entropy of the reduced state defines the block entanglement entropy quantifying the bipartite entanglement across the blocks separation boundary. Panel (d): From block entanglement to SE between subsystems. We illustrate the case of the long-distance SE and QCMIs $I_{(3)}(A:B|C)$ and $I_{(4)}(A:B|C_1)$ between two specific disconnected subsystems, the system edges $A$ and $B$ separated by the bulk $C=C_1C_2$. This is the case of interest, e.g., in the study of topological superconductivity in quantum matter. The double--headed arrows represent pictorially the spatial ranges of the different quantum correlation patterns.}
\label{Figure1}
\end{figure*}

The two inequivalent QCMIs $I_{(3)}$ and $I_{(4)}$ define two different upper bounds (corresponding, respectively, to a tripartite and a quadripartite multipartition of the system) to the same bipartite SE $E_{sq}(\rho_{AB})$ between subsystems $A$ and $B$ in the reduced state $\rho_{AB}$. It is immediate to verify that any further splitting of the reminder $C$ in multipartitions of higher order $C=C_1C_2C_3 \dots C_n$ is redundant in the sense that it defines a set of $n$ QCMIs of the form $I_{(4)}(A:B|C_i) \, , (i=1, ... , n)$. Taking into account the mathematical properties of the von Neumann entropies, including subadditivity and the triangle inequality, one has the chain of inequalities
\begin{equation} 
I_{(3)}(A:B|C) \geq I_{(4)}(A:B|C_1) \geq E_{sq}(\rho_{AB}) \, .
\label{inequality}
\end{equation}
The two fundamental constructions, respectively via the tripartition $ABC$ and the quadripartition $ABC_1C_2$, are summarized in panels (a) and (b) in Fig.~\ref{Figure1}.

The exact $A$--$B$ bipartite SE in state $\rho_{AB}$ is obtained by computing the infimum of the two expressions Eqs.~\ref{upper3}--\ref{upper4} over all extensions of unbounded dimension $\rho_{ABC}$ and $\rho_{ABC_1}$, respectively. This is in general an exceedingly hard task; on the other hand, the true SE is readily obtained whenever there are lower bounds available that coincide with some upper bounds, for instance like the ones provided by Eqs.~\ref{upper3}--\ref{upper4}.

Summarizing what has been discussed so far, we have shown how by resorting to the QCMI and the SE one moves from the elementary paradigm of bipartite pure--state block entanglement entropy defined over minimal, irreducible connected bipartitions of the global system to a general framework of bipartite entanglement between any two subsystems, either connected or disconnected, defined over arbitrary multipartitions of the global system. Moreover, we have introduced two classes of upper bounds to the exact bipartite SE that are defined, respectively, over tripartitions and quadripartitions of the global system. 

The transition from bipartitions to multipartitions and from block entanglement to SE between generic subsystems is represented pictorially in panels (c) and (d) of Fig.~\ref{Figure1}, where we illustrate schematically how such transition allows in principle to discriminate systems with broken symmetries and short-ranged entanglement from systems with a different type of global order and long-distance entanglement between disconnected subsystems and thus also between boundaries (edges) in the case of systems with open boundary conditions (open chains, open ladders, etc). In the next sections we investigate in detail some significant consequences of this paradigm shift in the study of topological quantum matter.

\section{Squashed entanglement and topological superconductors}
\label{topoentang}

\subsection{The quest for order parameters}

The archetypal model of topological $p$-wave superconductivity supporting edge Majorana zero energy modes (MZEMs) is the Kitaev Hamiltonian of spinless fermions on a 1D lattice \cite{Kitaev2001}. Proposals for realistic implementations of the Kitaev chain consider heterostructures made of semiconducting nanowires coupled to $s$-wave superconducting substrates \cite{Lutchyn2010,Oreg2010,Nadj2013,Lutchyn2018}. 

Experimental evidence of MZEMs localized at the system edges has been obtained in the study of the tunnel conductance of an InAs nanowire proximized by an $s$-wave superconductor \cite{Mourik2012} and in the scanning tunneling microscopy of iron-atom chains deposited on lead substrates \cite{Nadj2014}.

The one-dimensional Kitaev model can be generalized to geometries of significantly increasing complexity, e.g. via Kitaev ladders and ties, to describe coupled superconducting nanowires with a phase diagram hosting a rich variety of different topological phases \cite{Potter2010,Zhou2011,Wakatsuki2014,Schrade2017,Maiellaro2018,Maiellaro2020}.

In view of its conceptual significance and potential realizability in realistic systems of condensed matter physics, the Kitaev model has become a central paradigm in the study of topologically ordered phases of matter hosting MZEMs and, more generically, edge modes and edge states. 

The problem of identifying unambiguous signatures of topologically ordered phases, such as topological invariants and/or nonlocal order parameters, has turned out to be a highly nontrivial task addressed in a number of different approaches. One can look to momentum-space properties \cite{Shapiro2020}, when, after imposing periodic boundary conditions, the translational invariance is restored. This meaningful relation between the edges and the bulk of a system is also known as bulk-edge correspondence \cite{Shapiro2020} and leads to the definitions of geometric indices $Q$ which signal the presence and the number of topological zero-energy states of matter. 

Unfortunately, for disordered systems or, in general, when translational invariance cannot be restored, or also in the presence of interactions, the topological invariants $Q$ are rarely readily accessible and hence useful both analytically and experimentally~\cite{Hegde2016,Katsura2018,DeGottardi2011,Zhan2013,PhysRevB.31.3372}. 

In a more fundamental approach one tries then to identify topological invariants based on the patterns of nonlocal quantum correlations and the entanglement properties of the system, such as the bipartite entanglement spectrum \cite{Pollmann2010,Fidkowski2010,Turner2011}.

Indeed, the entanglement spectrum becomes two-fold degenerate when a quantum system undergoes a topological phase transition \cite{Dalmonte2013,Pollmann2021,Stoudenmire2011}; however, this is not a discriminating signature of topological order: exactly the same degeneracy is featured by any system with the same Hamiltonian symmetries that can support symmetry-breaking order \cite{Pollmann2010,Franchini2017}. As a consequence, block entanglement based on simple bipartitions of the system into two blocks does not provide the information necessary to identify and discriminate unambiguously topological order associated to edge modes and edge states. 

Suitable \emph{ad hoc} combinations of R\'enyi entropies of disconnected and partially overlapping multipartitions actually allow to define quantized topological invariants able to detect topologically ordered phases in one-dimensional topological superconductors~\cite{Dalmonte2020}. Unfortunately, these invariants are not entanglement monotones between distinguishable subsystems, let alone entanglement measures, and thus are devoid of any clear meaning in terms of nonlocal quantum correlations. Because of their \emph{ad hoc} nature and lack of conceptual significance and physical motivation, one has no clue on how to introduce such invariants systematically for the characterization of different types of topologically ordered phases or to extend them to general situations such as finite temperature and/or complex geometries and higher dimensions.

In the present work, we discuss SE in the context of topological quantum matter featuring edge modes and edge states. We show that the edge-to-edge, long-distance bipartite SE between the (disconnected) edges of one-dimensional topological systems is a (nonlocal) order parameter, properly quantized, featuring the correct scaling behavior at the critical point, characterizing the ordered phases of topological superconductors in various one-dimensional and quasi one-dimensional geometries, and discriminating them unambiguously from other classes of ordered phases of matter.

Shifting focus from block entanglement entropy to SE between disconnected subsystems separated by arbitrary distances finds a basic physical motivation in the observation that systems with topological order display bulk band gaps like those of ordinary insulators and conducting surface states that are topologically protected by some symmetries. This naturally prompts to look for nonlocal correlations between subsystems (e.g. the system edges) rather than the block entanglement between two halves of the total system. In turn, this implies replacing bipartitions with multipartitions, and resorting to SE as the prospective quantity able to:

\begin{itemize}

\item Quantify the {\it prima facie} bipartite, long-distance edge-edge entanglement in the presence of edge modes.

\item Discriminate unambiguously topological order from symmetry-breaking order and distinguish between different classes of topologically ordered phases.

\end{itemize}

In the following we will consider one-dimensional systems hosting MZEMs and for such systems we will investigate the two bipartite edge-to-edge QCMIs, $I_{(3)}(A:B|C)$ and $I_{(4)}(A:B|C_1)$ that realize two upper bounds on the true SE $E_{sq}(\rho_{AB})$ between edge $A$ and edge $B$. In particular, we will show that the QCMI $I_{(4)}$ discriminates the one-dimensional Kitaev model featuring edge modes and long-distance entanglement between the edges from the Ising chain, its symmetry-breaking counterpart featuring no edge modes and a short-distance entanglement structure. 

For the Kitaev chain we will show that $I_{(3)}$ and $I_{(4)}$ coincide and satisfy all the requirements for a genuine topological order parameter, including: quantization as topological invariant at the exact topological ground-state degeneracy point and throughout the topologically ordered phase; scaling with the system size at the phase transition point; stability with respect to interactions; and robustness against localized disorder and imperfections. 

Moreover, for the Kitaev chain the topological invariant in fact coincides with a lower bound to the SE, and therefore the QCMIs $I_{(3)}$ and $I_{(4)}$ indeed coincide with the long-distance, bipartite SE $E_{sq}$ between the chain edges at the exact topological degeneracy point and throughout the entire topologically ordered phase. The transition from simple bipartitions and block entanglement entropy to multipartitions and edge to edge long-distance entanglement is illustrated pictorially in panels (c) and (d) of Fig.~\ref{Figure1}.

We will then generalize the method to consider models of topological superconductors defined on geometries of higher complexity, the quasi one-dimensional Kitaev ladder \cite{Maiellaro2018} and the Kitaev tie \cite{Maiellaro2020}. The Kitaev ladder model is obtained by coupling two Kitaev chains by means of transverse hopping and pairing terms. In fact, the Kitaev chain and the Kitaev ladder belong to the two different topological classes $D$ and $BDI$ which are characterized by two different topological bulk invariants, respectively the Pfaffian invariant \cite{Budich2013} and the winding number~\cite{Maiellaro2018}. In the case of the Kitaev tie, the long-range hopping term added to the Hamiltonian of the Kitaev chain define a knotted--ring geometry which can be rearranged in the form of a tie. The model is then the simplest realization of geometric frustration with no associated bulk~\cite{Maiellaro2021}.

For all three models SE identifies and characterizes the different topological behaviors of the systems, well characterized by the band topology for the Kitaev chain and the Kitaev ladder and by the Majorana polarization and topological transfer matrix for the Kitaev tie~\cite{Maiellaro2021}. We will show that the QCMI $I_{(4)}$ identifies the topological phase transitions in the Kitaev ladder and distinguishes between the different topologically ordered phases of the model corresponding to different numbers of Majorana excitations. 

Remarkably, in the case of the Kitaev tie, a system which lacks a clearly identifiable bulk, the SE is still very efficient in characterizing the topological phases of the model even if, as expected, perfect quantization is partially blurred due to the absence of a clear physical bulk-boundary separation.

The theoretical framework based on multipartitions, bipartite QCMIs and bipartite SE between generic subsystems (either connected or disconnected) can be generalized to finite temperature and non-equilibrium, to multipartite entanglement, and to many-body systems defined in higher-dimensional geometries and with different classes of boundary conditions. In the conclusions we will discuss some basic aspects of these future generalizations and how, for each case study, they will depend crucially on the localization and connection properties of edge modes and edge states.

\subsection{Topological squashed entanglement}

For the model systems defined in the following we consider tripartitions and quadripartitions as sketched in Fig.~\ref{Figure1} and we proceed to compute the QCMI upper bounds $I_{(3)}$ and $I_{(4)}$ on the SE $E_{sq}(\rho_{AB})$ between the edges $A$ and $B$ given a bulk $C=C_1C_2$. 

In a condensed matter setting $I_{(3)}$ and $I_{(4)}$ are defined in terms of the many body ground state, i.e. they are the QCMIs of the edges $A$ and $B$ conditioned, respectively, on the existence of the total bulk $C$ and the partial bulk $C_1$. When going to sufficiently large system sizes such that the edges are sufficiently distant from each other and fully decoupled from the bulk, a nonvanishing $E_{sq}(\rho_{AB})$ establishes the existence of a topological squashed entanglement (TSE), i.e. a nontrivial long-distance quantum correlation between the edges. 

Although the physical mechanisms are rather different, the edge-edge fermionic TSE is reminiscent of other forms of long-distance entanglement (LDE) that are established by entanglement monogamy between the end points of dimerized and quasi-dimerized spin-$1/2$ chains with diverse patterns of nearest-neighbor couplings or patterns of competing finite-range interactions~\cite{Giampaolo2007,Giampaolo2010,Giampaolo2015}. Specifically, in these previous works we have shown that systems defined on $1D$ chains with open boundary conditions will exhibit a nonvanishing end-to-end, long-distance entanglement (LDE) whenever: $I)$ there is a weak coupling between the edge regions and the bulk, or $II)$ there is a pattern of alternating weak and strong nearest-neighbor couplings that leads to an effective dimerization of the system. In turn, these two instances can be seen as particular cases of more general patterns of modular entanglement~\cite{Gualdi2011} and surface entanglement on networks~\cite{Zippilli2013}. 

Instance $I)$ yields an LDE between the chain ends that tends to decay very slowly with the distance, while instance $II)$ realizes a perfect LDE between the chain ends that does not decay with and is independent of the size of the chain. The further form of LDE that we have now discovered is 
$III)$ the end-to-end TSE in the topological phase of one-dimensional and quasi one-dimensional fermionic systems. It is unclear at the moment whether forms $I)$ and $II)$ of LDE and form $III)$ are related and share some common feature/origin. We plan to investigate the possible relations, also in connection with the intriguing possibility that complex patterns of interaction strengths might induce a kind of topological order also in some classes of spin-$1/2$ systems.

\begin{figure}
\includegraphics[scale=0.125]{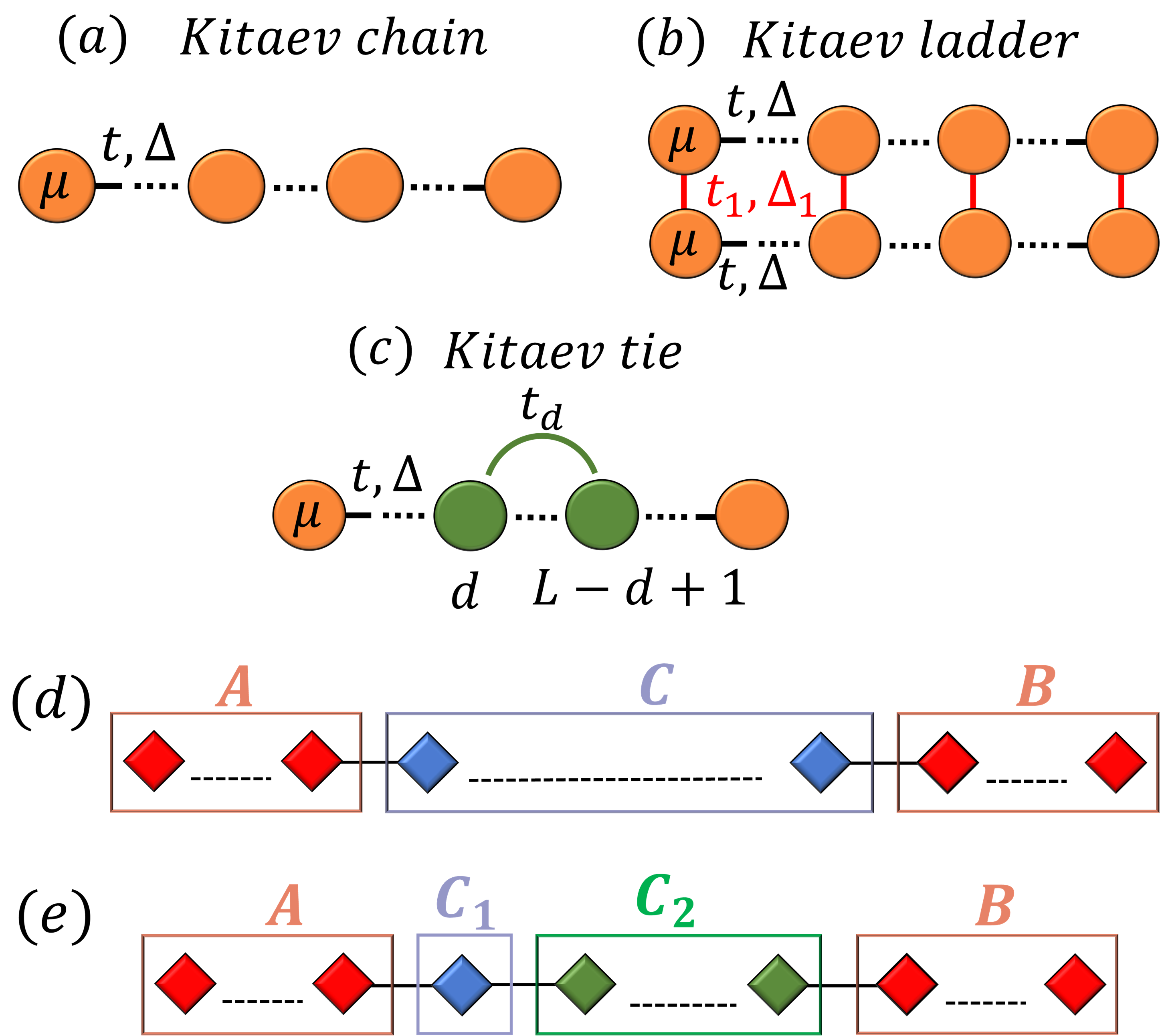}
\caption{Schematic of a Kitaev chain (a), a two--leg Kitaev ladder (b) and a Kitaev tie (c). The Kitaev ladder is obtained coupling two Kitaev chains by superconducting and hopping terms. For a Kitaev chain of length $L$, the Kitaev tie is obtained by adding a long--range hopping term coupling a site at position $d$ with the symmetric one at the position $L-d+1$. The basic tripartition and quadripartition are reported, respectively, in panels (d) and (e). In both panels the edges are denoted by $A$ and $B$ (red color); in panel (d) the total bulk $C=C_1C_2$ is reported in blue; in panel (e) the bulk portions $C_1$ and $C_2$ are reported in blue and in green, respectively. The diamonds are generic and can correspond to a single fermionic site when referred to a chain or a tie and to two fermionic sites when referred to a ladder. In the quadripartition, when not otherwise specified, the length $L_{C_1}$ of the bulk portion $C_1$ is fixed at $L_{C_1}=1$.}
\label{Figure2}
\end{figure}

In panels (a)--(c) of Fig.~\ref{Figure2} we provide a sketch of the three model geometries considered (Kitaev chain, Kitaev ladder, Kitaev tie), while in panels (d) and (e) we draw a synthetic scheme of how the multipartitions reported in Fig.~\ref{Figure1} are applied to these three explicit cases. 

As detailed in the next sections, both the Kitaev ladder and the Kitaev tie can be obtained, respectively, from two coupled Kitaev chains and by adding to the Hamiltonian of a single chain a symmetric long--range hopping that couples a single lattice site at position $d$ with its symmetric counterpart at position $L - d + 1$. In panels (d) and (e) of Fig.~\ref{Figure2} we provide a schematics of the two basic multipartitions for the model systems considered.

In the following, the lengths of the various subsystems will be denoted by $L_{\alpha}$, with $\alpha=A$, $B$, $C$, $C_1$, $C_2$. Throughout, the lengths of the two edges are assumed to coincide, $L_A=L_B$. Denoting by $L$ the chain length, when not otherwise specified we will set $L_A = L_B = L_e$, with $L_e = L/3$. The latter choice allows to take into account the exponential bulk--edge decoupling behavior as a function of the system size that is typical of topological modes. 

Moreover, as explained later on (see Fig.~\ref{Figure5}, panel (a)), our results will be independent of the relative size of the two sub-bulks, so that without loss of generality we will take $L_{C_1}=1$ throughout, as reported in panel (e) of Fig.~\ref{Figure2}.  

A crucial step in order to compute the TSE consists in diagonalizing the reduced density matrices of the various subsystems, once the many body ground state density matrix $\rho= |\Psi \rangle_{G} {}_{G}\langle \Psi |$ is assigned. 

When interactions are neglected, the Kitaev-type Hamiltonians are quadratic in the fermionic degrees of freedoms and can be diagonalized by a Bogoliubov transformation. In this case, an appropriate approach to compute the von Neumann entropies of the subsystems has been introduced by Peschel~\cite{Peschel_2009}. 

The Bogoliubov transformation ensures a direct access to fermionic correlations on the many body ground state $|\Psi \rangle_{G}$ and the reduced density matrix of a given subsystem $\alpha$ can be recast in the form of a thermal density matrix, $\rho_{\alpha}=(1/Z)e^{-\mathcal{H}_{\alpha}}$, where $\mathcal{H}_{\alpha}=\sum_{l=1}^{L_{\alpha}} \epsilon_l f^\dagger_l f_l$ is the effective entanglement Hamiltonian of the reduced system~\cite{Eisler_2018,PhysRevB.62.4191,PhysRevLett.121.200602,PhysRevB.64.064412}, and the constant factor $Z$ (the "partition function") ensures the correct normalization $Tr(\rho_{\alpha})=1$. 

Given the lattice fermionic creation and annihilation operators $\{c_i, c^\dagger_i \}$, the spectrum of the entanglement Hamiltonian can be evaluated numerically by solving the eigenvalue problem:
\begin{equation}
 	\big( 2T - 1 - 2D \big) \big( 2T - 1 +2D \big) \phi_l=\tanh^2 \bigg(\frac{\epsilon_l}{2} \bigg) \phi_l \, ,
\label{Peschel1}
\end{equation}
where $T_{ij} = \langle \Psi | c^\dagger_i c_j |\Psi \rangle$ and $D_{ij}=\langle \Psi|c^\dagger_i c^\dagger_j |\Psi \rangle$ are the matrix elements of the fermionic two-point correlations in a quantum state $|\Psi \rangle$ (for instance, the ground state $|\Psi \rangle_G$), and $1$ is the identity matrix. The index $l$ runs over all the lattice sites belonging to subsystem $\alpha$. 

Due to the form of $\rho_{\alpha}$, the von Neumann entropy of the reduced state $S(\rho_{\alpha})$ can be obtained in terms of the eigenvalues of entanglement Hamiltonian:
\begin{equation}
	S(\rho_{\alpha}) = \sum_l \ln(1+e^{-\epsilon_l})+\sum_l \frac{\epsilon_l}{e^{\epsilon_l}+1}.
\label{Peschel2}
\end{equation}
The Peschel algorithm for free fermionic systems reduces the computational efforts and allows to access the entanglement properties for systems of very large sizes. 

When interactions are included in the models, the mapping to the effective entanglement Hamiltonian and the Gaussian thermal structure of the reduced density matrix are no longer applicable and one must resort to numerical techniques, e.g. the density matrix renormalization group (DMRG) method, particularly well suited to treat one-dimensional systems~\cite{PhysRevLett.119.246401,ORUS2014117,SCHOLLWOCK201196}. 

Complementary to numerical methods, one can resort to direct numerical matrix diagonalization in the parameter region that allows for exact analytic expressions of the state vectors, such as along the exact factorization lines of interacting spin models and Kitaev--type fermionic systems~\cite{Illuminati2008,Illuminati2009,Illuminati2010,Katsura2015}, as we will discuss in the following. 

\section{Topological squashed entanglement of the Kitaev chain}
\label{KC}

\subsection{The nonlocal order parameter}
Here and in the following we investigate the paradigmatic case of the fermionic Kitaev chain~\cite{Kitaev2001}, considering first the non-interacting model:
\begin{equation}
	H_K=\sum_{j=1}^L-\mu c^\dagger_{j}c_{j}+\sum_{j=1}^{L-1}(\Delta c^\dagger_{j+1}c^\dagger_{j}-tc^\dagger_{j}c_{j+1}+h.c.).
	\label{Kitaev}
\end{equation}

The Hamiltonian $H_K$ in Eq.~\ref{Kitaev} describes a system of spinless fermions confined to a one dimensional lattice of length $L$ with on site creation and annihilation operators $c_j^\dagger$, $c_j$ ($j=1, \dots, L$), subject to a $p$-wave superconducting coupling. The coefficients $\Delta$, $t$ and $\mu$ are respectively the strength of the superconducting pairing, the nearest-neighbor hopping amplitude and the on-site energy offset. 

This model features a topological phase for $\mu < 2t$ and $\Delta \neq 0$ with two robust Majorana zero-energy modes localized at the two edges. The energy $E_M$ of such MZEMs scales exponentially to zero with the system's size and is expected to vanish asymptotically in the thermodynamic limit. 

\begin{figure*}
\includegraphics[scale=0.145]{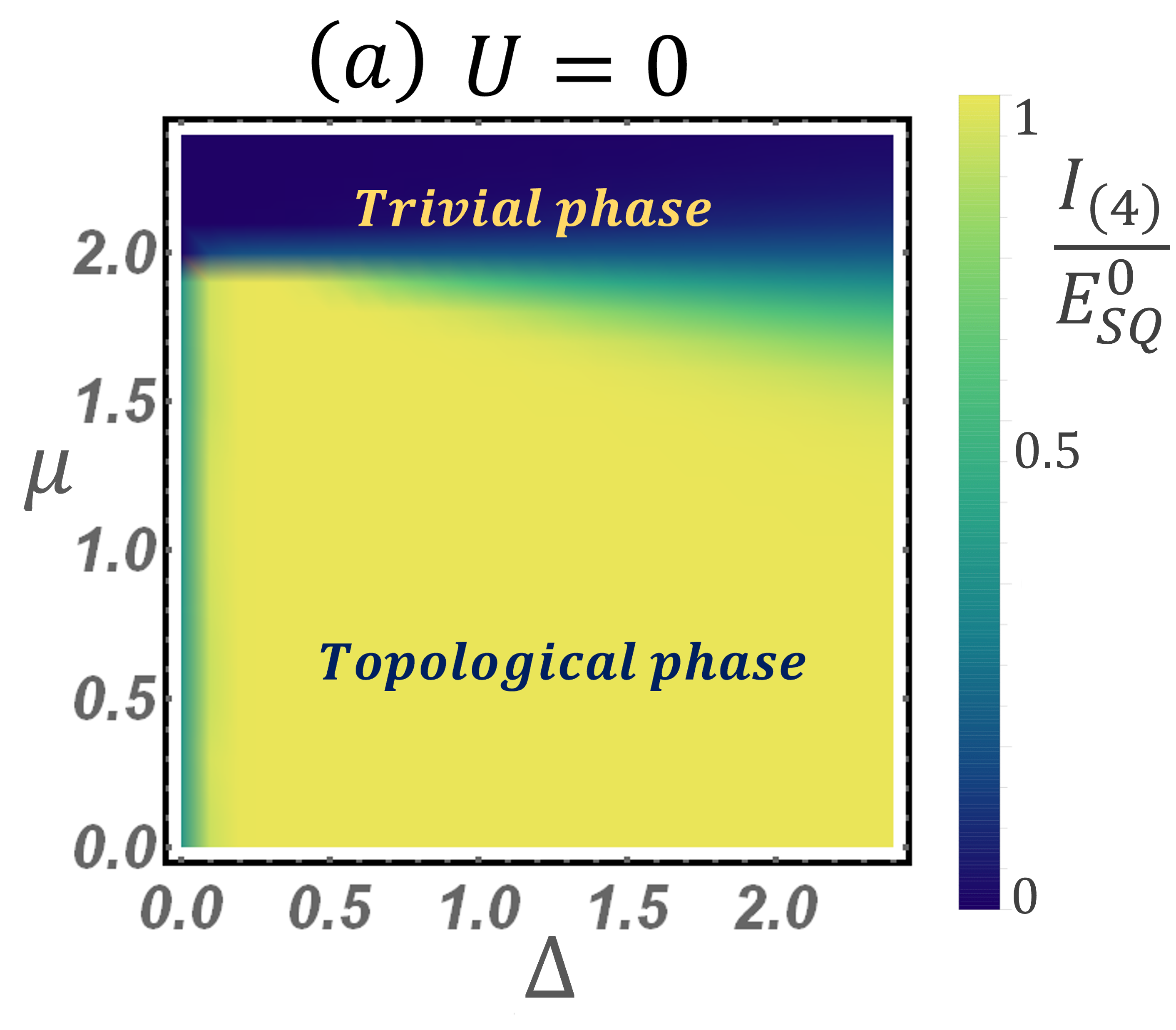}
\hspace{1cm}
\includegraphics[scale=0.13]{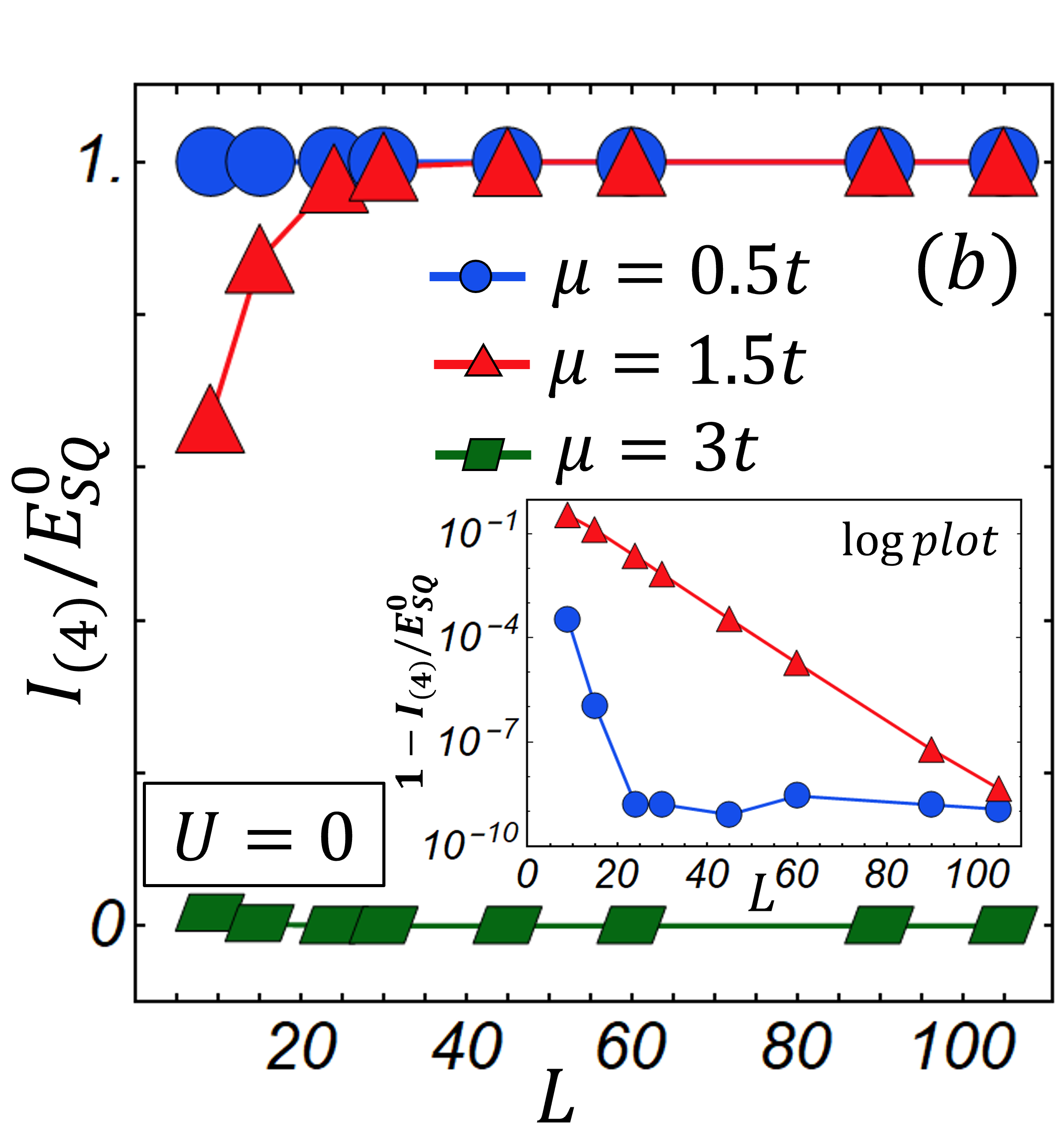}
\hspace{1.3cm}
\includegraphics[scale=0.14]{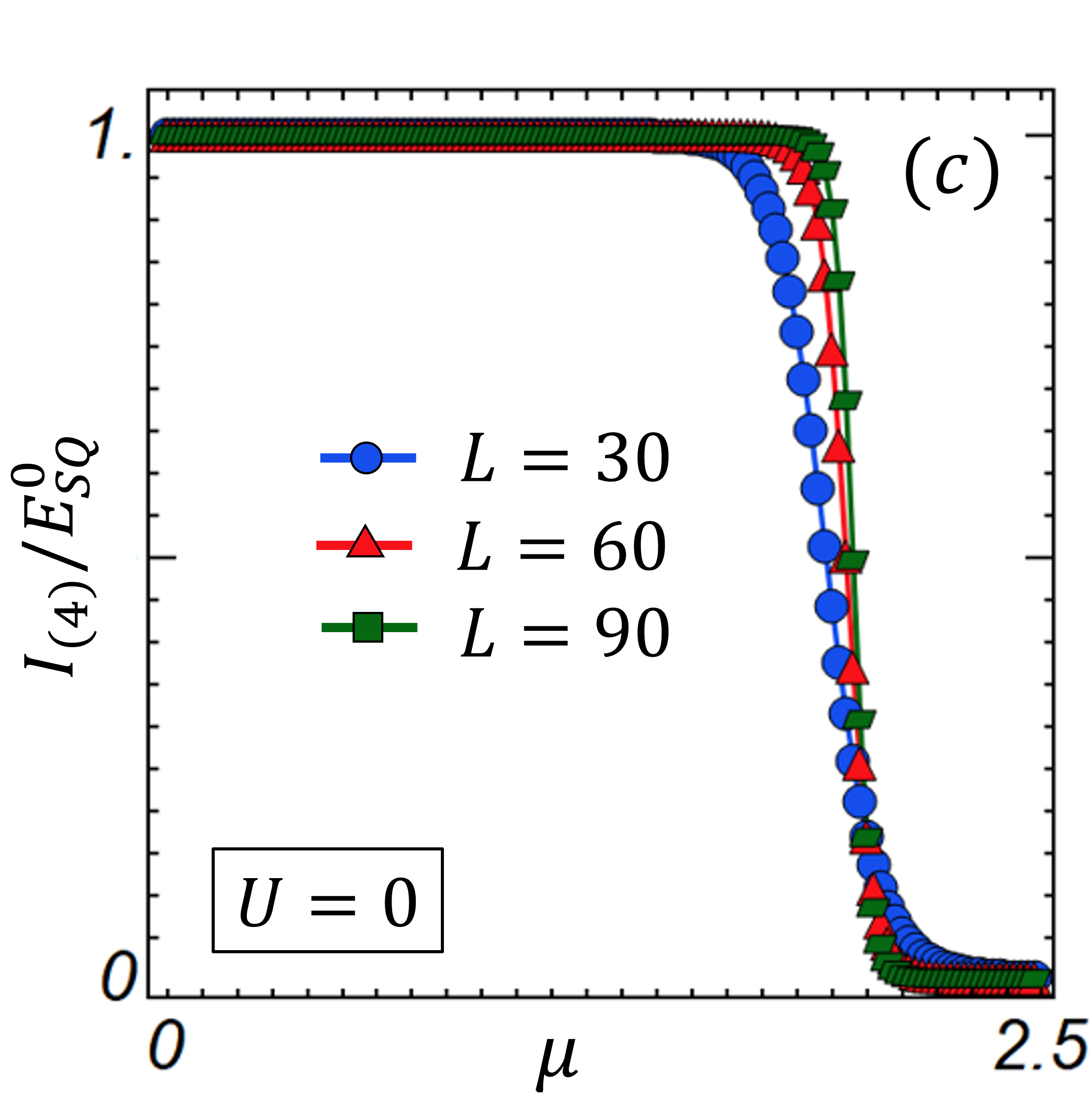}
\caption{Panel (a): Phase diagram of the non-interacting Kitaev chain as measured by the two-dimensional contour plot of the ratio between the QCMI $I_{(4)}$ and the quantized TSE $E^0_{sq} = \log (2)/2$ on the $\Delta$--$\mu$ plane with a grid of $25 \times 25$ points, reference hopping amplitude $t=1$, chain length $L=60$, and edge length $L_e=L/3$. Panel (b): Scaling-size effects on $I_{(4)}/E^0_{sq}$ as a function of $L$ for different values of the chemical potential $\mu$. Panel (c): Scaling of $I_{(4)}/E^0_{sq}$ as a function of $\mu$ for different values of $L$.}
\label{Figure3}
\end{figure*}

In the limit of small on site energy offset (chemical potential), precisely at the analytically solvable point $\mu=0$ and $t=\Delta$, the model features exact two-fold topological ground-state degeneracy and two MZEMs with exactly $E_M = 0$, independently of the chain length. The remaining non-topological modes of the spectrum are gapped and form a band. In the opposite regime of large on site energy offset $\mu>2t$ the Kitaev wire is a trivial band insulator. 

By construction the Hamiltonian satisfies the particle-hole symmetry and belongs to the class $\mathcal{D}$ of the ten-fold classification \cite{Altland1997}, with topological invariant given by the Pfaffian invariant \cite{Budich2013}.

For the non-interacting Kitaev chain, we have computed the two QCMIs $I_{(3)}$ and $I_{(4)}$ and verified that $I_{(3)} = I_{(4)}$ throughout the entire phase diagram, so that in the reminder of the present subsection we will focus only on $I_{(4)}$. The physical origin of the equality between the two QCMIs in the topological case will become clear in the following.

In panel (a) of Fig. \ref{Figure3} we report the phase diagram obtained by means of the ratio between the QCMI $I_{(4)}$ and the constant quantity $E^0_{sq} = \log (2)/2$. One finds that throughout the entire topological phase the ratio $I_{(4)}/E^0_{sq} = 1$ so that $I_{(4)}$ is quantized exactly at the value $I_{(4)} = E^0_{sq} = \log (2)/2$. Moreover, one has that $I_{(4)} = 0$ throughout the entire trivial phase. 

Finite-size effects are clearly visible near the phase transition line at $\mu=2 t$, as also shown in panel (b) of Fig.~\ref{Figure3}, where the ratio $I_{(4)}/E^0_{sq}$ is plotted as a function of the length $L$ of the chain for $t=\Delta=1$ and three different values of the chemical potential $\mu$, respectively close to the exact topological point, in the topological phase close to the critical point, and well into the trivial phase. 

In particular, $I_{(4)}$ scales exponentially to the quantized value $E^0_{sq}$ inside the topological phase (red curve) and it coincides with $E^0_{sq}$ independently of the size of the chain for values of $\mu$ sufficiently close to the analytic point (blue curve). The QCMIs vanish identically at all values of $\mu$ in the trivial phase (green curve). 

Finally, in panel (c) of Fig.~\ref{Figure3} we report the behavior of the ratio $I_{(4)}/E^0_{sq}$ as a function of $\mu$ for different values of the length of the chain $L$, showing the correct approach to quantization inside the topological phase and the correct scaling behavior approaching the critical point. 
These results show that the QCMIs $I_{(3)}$ and $I_{(4)}$ provide a completely equivalent detection and characterization of the topologically non-trivial regime of a superconducting wire. 

In fact, by applying the lower-bound inequalities holding on SE~\cite{CarlenLieb2012,LiWinter2014,Brandao2015,FawziRenner2015}, it turns out that the constant value $E^0_{sq} = \log (2)/2$ is a lower bound to the true SE $E_{sq}(\rho_{AB})$ between the edges at the exact topological point $\mu = 0$. By continuity, the bound extends to neighboring values of $\mu$. 
Since the upper bound $I_{(4)}$ and the quantized lower bound $E^0_{sq}$ coincide in this region, both coincide with the true long-distance SE $E_{sq}(\rho_{AB})$ between the two disconnected edges $A$ and $B$. 

In conclusion, collecting all the above results, the quantized TSE 
\begin{equation}
E_{sq}(\rho_{AB}) = E^0_{sq} = \log ( \sqrt{2} ) = \frac{\log (2)}{2} 
\label{TSE}
\end{equation}
is the nonlocal order parameter for a Kitaev superconducting wire. 

The physical origin of a quantized non-vanishing long-distance entanglement between the system edges in the topological phase arises from the interplay between the exponential bulk-edge separation due to the topological nature of the system and the fundamental monogamy property of nonlocal quantum correlations~\cite{Koashi2004}.

As the two edges progressively decouple from the bulk their mutual information and correlation are enhanced by the monogamy constraint on the shared information and on the amount of shareable entanglement among the different subsystems. When the system moves away from the topological phase, correlations become once again short-ranged and the long-distance entanglement between the two extremes of the chain collapses.

The quantized value of the edge--edge TSE is $\log(2)/2$ rather than the maximal Bell--pair entanglement $\log(2)$. This fact shows that the TSE is fully sensitive to and takes correctly into account the nature of the topological modes, distinguishing between the entanglement of half-fermions, like MZMEs, from the maximal Bell--pair entanglement.

The TSE can be compared with the so-called entanglement topological invariant (ETI), a nonlocal order parameter for one-dimensional topological superconductors introduced in Ref. \cite{Dalmonte2020}. This quantity is defined in terms of a combination of reduced von Neumann entropies associated to a pseudo-quadripartition obtained by considering two disconnected and partially overlapping sets, $A_1$ and $A_2$, their union $A_1 \cup A_2$, and their intersection $A_1 \cap A_2$. As such, the ETI has no sensible relation to a real physical multipartition of the system and the corresponding nonlocal quantum correlations between separated and non-overlapping subsystems. In particular, it is not an entanglement monotone, let alone a genuine entanglement measure between any two separated and distinguishable subsystems like, e.g., the system edges.

One particularly limiting tenet of the ETI as introduced in Ref. \cite{Dalmonte2020} is that a key ingredient for its derivation is the existence of disconnected partitions of the system. In fact, we have just shown and explained in detail why a fully connected edge-bulk-edge system tripartition and the associated QCMI $I_{(3)}$ yield exactly the same characterization of the edge modes and the topologically ordered phases as the one provided by the quadripartition edge-bulk-bulk-edge and the associated QCMI $I_{(4)}$.

Also, in Ref. \cite{Dalmonte2020} it is found that the ETI takes the value $\log (2)$, i.e. twice the value of the TSE. As discussed above, the value $\log (2)/2$ assumed by the TSE is the one that takes correctly into account the statistics of the Majorana excitations versus that of full Dirac fermions; moreover, the factor $1/2$ provides the correct reduction on pure states to the unique entanglement measure, i.e. the von Neumann entanglement entropy. Such correct identifications are a direct consequence of the precise physical identification of the system multipartition and general definition and properties of the edge-edge QCMI and TSE. No such comprehensive framework exists for the ETI. In conclusion, the scheme based on the edge-edge bipartite QCMI and TSE defined on multipartitions provides a well-defined, general physical framework to determine the correct nonlocal order parameter for topologically ordered phases in one dimension. Moreover, this framework can be extended in principle to the study of topological systems in any dimension and at any temperature, either at equilibrium or in non-equilibrium.

Concerning the experimental accessibility of topological squashed entanglement, the problem boils down to that of measuring quantum entropies of a set of reduced states in quantum many--body systems. A recent proposal relies on the thermodynamic study of the entanglement Hamiltonian for the direct experimental probing of von Neumann entropies via quantum quenches \cite{Mendes_Santos_2020}. Another possibility, specific for systems featuring topological order, consists in identifying minimum entropy states and then experimentally simulating the behavior of the associated von Neumann entropies via the classical microwave analogs of such states. In this way it is in principle possible to simulate various non-trivial instances of reduced entropies and topological order, as shown explicitly for the transition from a trivial phase to a $\mathcal{Z}_2$--symmetric topological phase \cite{Zhang_2019}.

Another intriguing possibility arises from the observation that highly informative bounds on von Neumann entropies, quantum conditional mutual information, and squashed entanglement can be constructed in terms of R\'enyi entropies~\cite{Brandao2015,FawziRenner2015}. A possible strategy is then to adapt to fermionic systems~\cite{Cornfeld2019} the schemes previously proposed for the experimental measurement of R\'enyi entropies in bosonic and spin systems~\cite{Abanin2012,Daley2012,Elben2018} and the corresponding experimental techniques that led to the first experimental measurement of the 2-R\'enyi entropy in a many-body system~\cite{Islam2015}. This possibility has been already suggested in Ref. \cite{Dalmonte2020} for the experimental measurement of the ETI.



\subsection{SE and TSE: discriminating between symmetry breaking magnets and topological superconductors}

A core question concerns the ability of a proposed nonlocal order parameter to discriminate topological superconductors from systems featuring local order parameters, spontaneous symmetry breaking and Ginzburg-Landau type of ordered phases, addressing their different entanglement properties and patterns. 

We focus our attention on the comparison between the spin-$1/2$ Ising chain and the fermionic Kitaev chain. Setting $t=\Delta=1$ and applying the global Jordan-Wigner transformations~\cite{Franchini2017}, the Hamiltonian in Eq. (\ref{Kitaev}) can be mapped into that of the $1D$ Ising model in transverse field: 
\begin{equation}
H_I = J \sum_{j=1}^{L-1} \sigma_j^x \sigma_{j+1}^x+h \sum_{j=1}^{L} \sigma_j^z ,
\label{Ising}
\end{equation}
with $J=-t$ and $h=\mu/2$. Both models share a two-fold degenerate ground state and the parity symmetry $\mathcal{Z}_2$, but the physics of the order displayed is radically different. While in the Ising model the $\mathcal{Z}_2$ spin reflection symmetry is spontaneously broken with the appearance of a nonvanishing local order parameter, the topological degeneracy in the Kitaev chain is realized with perfect conservation of the Hamiltonian symmetry and no local order parameters. 

Despite the different type of order that they support, the two models share the same bipartite entanglement spectrum and bipartite von Neumann block entanglement entropy. This is not surprising, since the nonlocal Jordan-Wigner transformation (a particular case of the general Klein transformation in quantum field theory) does not affect the short-distance entanglement structure across the separation between the system's halves for Hamiltonians sums of local (nearest-neighbor or short-ranged) interaction terms. 

The landscape is quite different when dealing with the structure of long-distance entanglement (LDE) between the end boundaries of a many-body system: clearly, we expect LDE to be significantly affected by global transformations acting on the entire system.

\begin{figure*}
	\includegraphics[width=6.9cm]{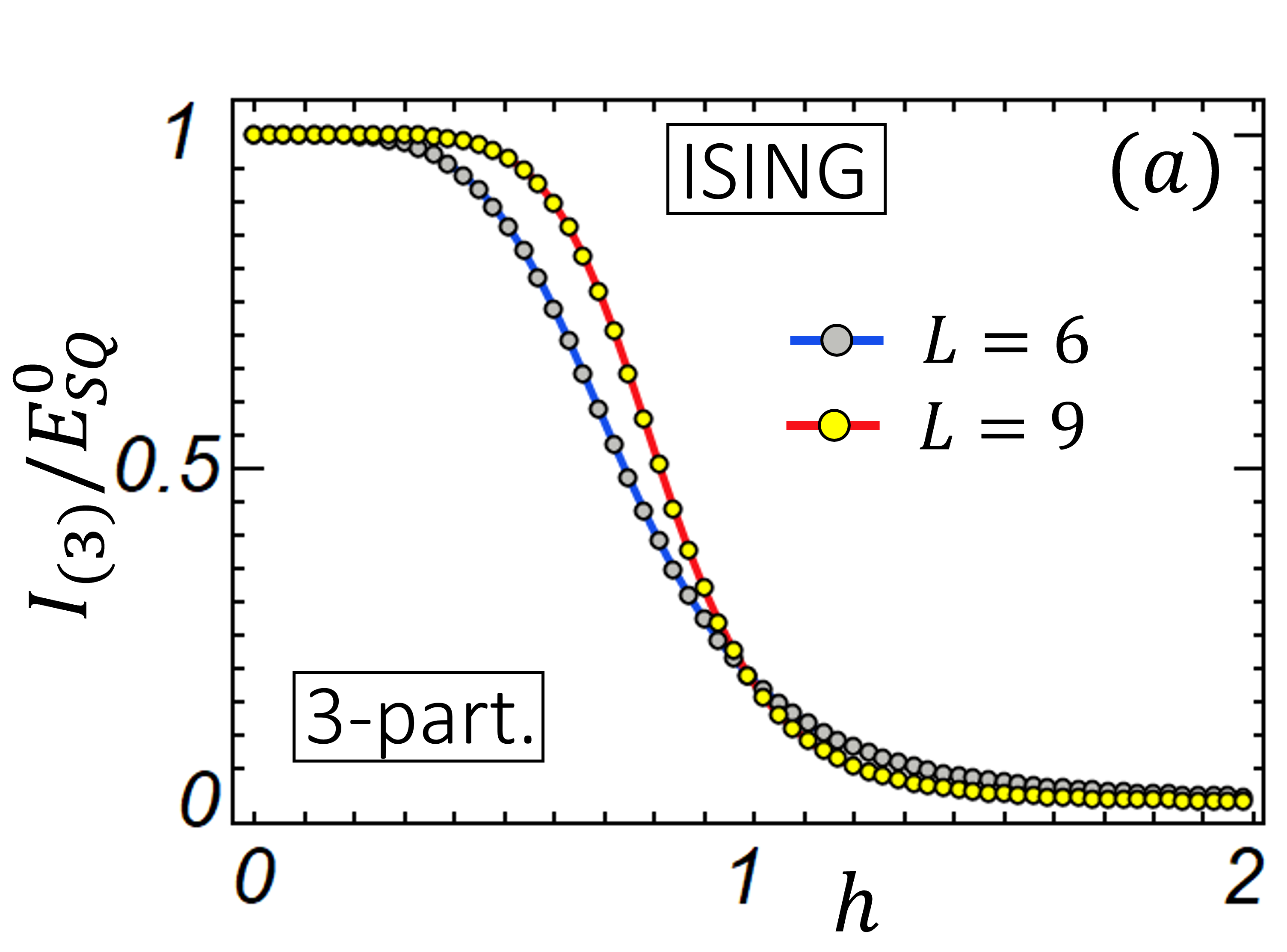}
	\hspace{1.5cm}
	\includegraphics[width=6.9cm]{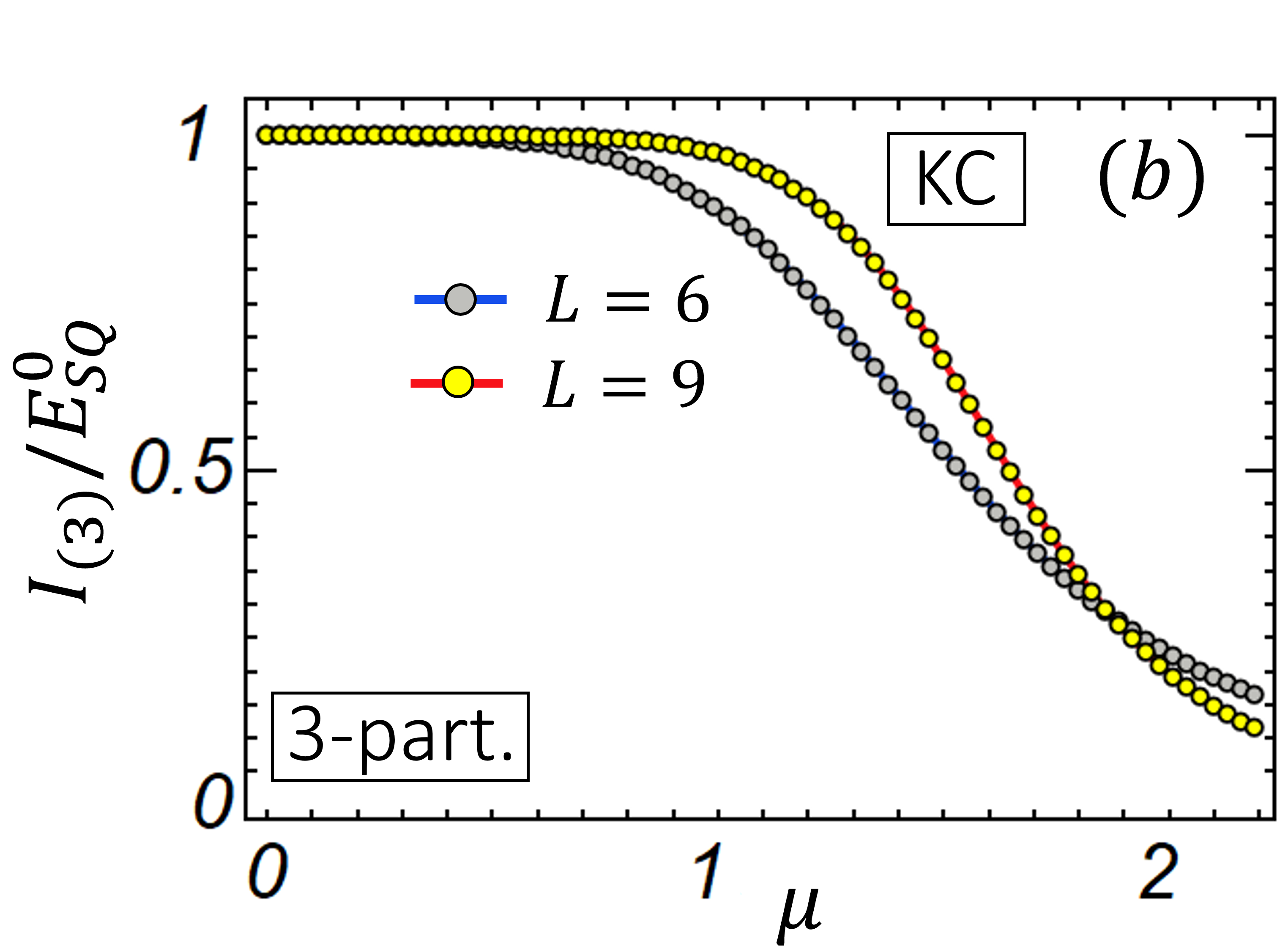}\\
	\vspace{0.3cm}
	\includegraphics[width=6.9cm]{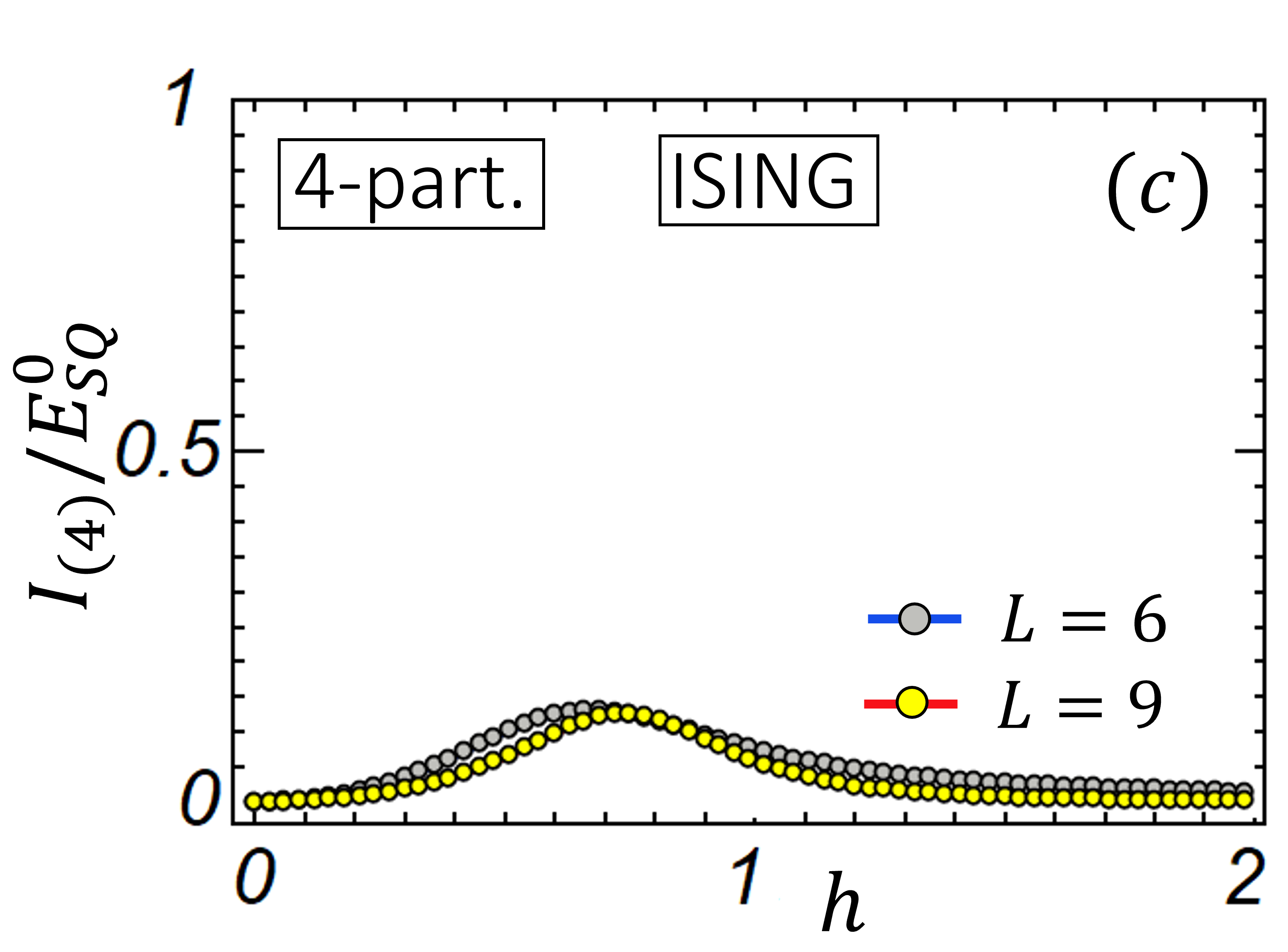}
	\hspace{1.5cm}
	\includegraphics[width=6.9cm]{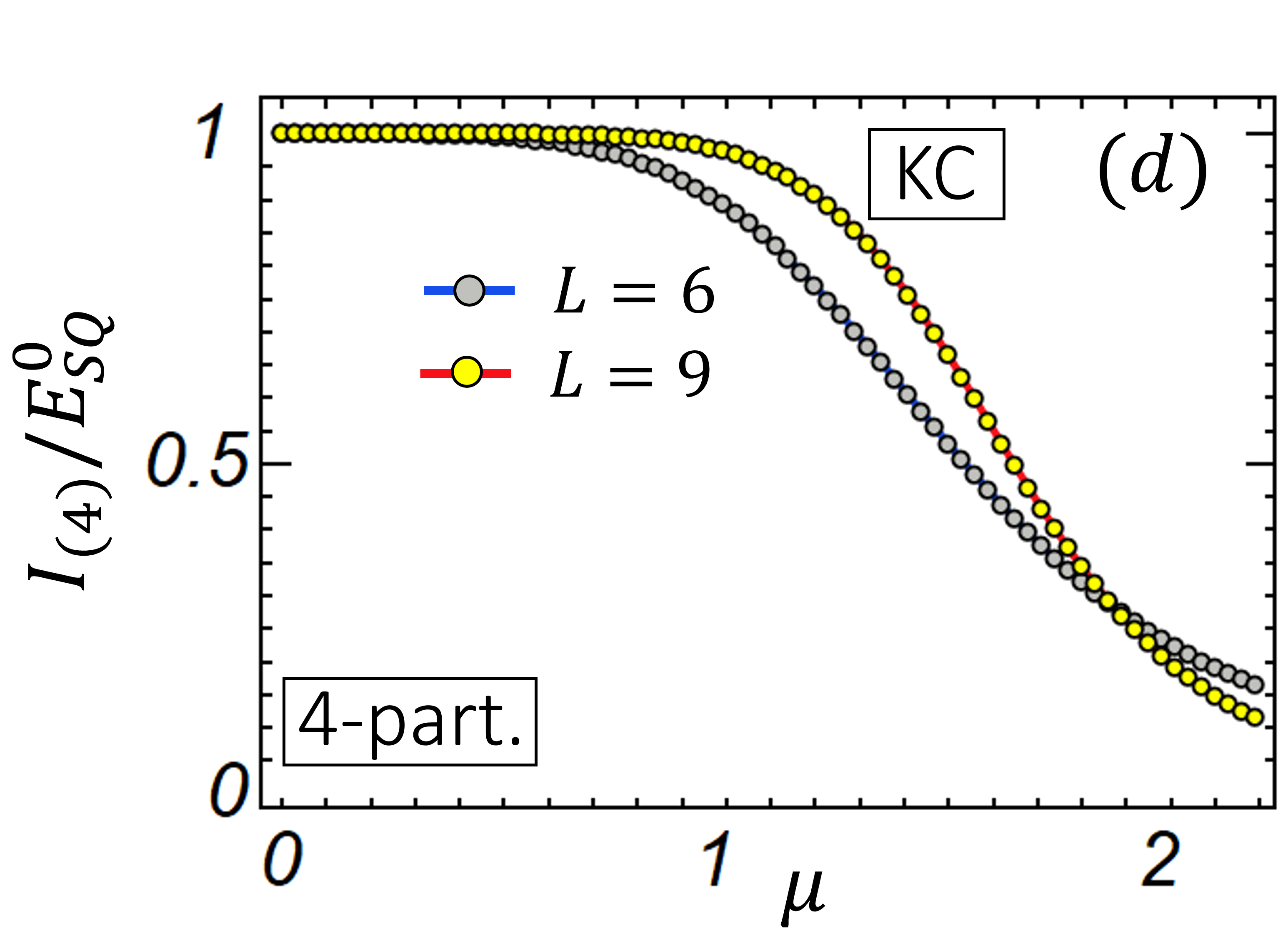}
	\caption{Ratios of the end-to-end QCMIs $I_{(3)}$ and $I_{(4)}$ to the quantized TSE unit $E^0_{SQ} = \log(2)/2$ for the Ising chain (panels (a) and (c)) as functions of the transverse field $h$, and for the Kitaev chain (panels (b) and (d)) as functions of the chemical potential $\mu$.  While for the Kitaev chain the equality $I_{(3)} = I_{(4)}$ always holds, for the Ising chain $I_{(3)} \neq I_{(4)}$ and $I_{(4)}$ tends to vanish as the system size increases. The QCMI $I_{(4)}$ thus discriminates between topological and symmetry-breaking systems. Throughout, the Hamiltonian parameters are set at $t=\Delta=1$ for the Kitaev chain and $J=-1$ for the Ising chain.}
	\label{Figure4}
\end{figure*}

In panels (a) and (b) of Fig.~\ref{Figure4} we report the behavior of the end-to-end QCMI $I_{(3)}$ for the Ising and the Kitaev chains for two lengths $L=6$ and $L=9$. The two curves intersect at the critical points $h=1$ and $\mu=2$. In the thermodynamic limit, the following behaviors are expected: 
$I_{(3),Ising}/E^{0}_{sq,Ising} = 1$ ($I_{(3),Kitaev}/E^{0}_{sq,Kitaev} = 1$) for $0<h<1$ ($0<\mu<2$) and $I_{(3),Ising}/E^{0}_{sq,Ising} = 0$ ($I_{(3),Kitaev}/E^{0}_{sq,Kitaev} = 0$) for $h>1$ ($\mu>2$). This analysis shows that the QCMI $I_{(3)}$ detects and characterizes both topological and symmetry breaking orders but is not capable of distinguishing between them.

In panels (c) and (d) of Fig.~\ref{Figure4} we report the behavior of the end-to-end QCMI $I_{(4)}$ for the Ising and Kitaev chains; we see that $I_{(4)}$ distinguishes topological superconducting phases from ordered phases with spontaneously broken symmetries.

For the Ising chain, we see that removing any finite part $C_2$ of the bulk $C=C_1C_2$ before performing the partial trace with respect to the reminder $C_1$ returns a strongly suppressed end-to-end QCMI $I_{(4)}$ that is expected to vanish in the thermodynamic limit. 

At variance with the symmetry-breaking case, the topology of the Kitaev chain is essentially concentrated on the boundaries, so that in the computation of the end-to-end QCMIs performing the partial trace with respect to the entire bulk $C$ is entirely equivalent to removing a part of the bulk ($C_2$) and then performing the partial trace with respect to the remaining part of the bulk ($C_1$). As a consequence, $I_{(3)} = I_{(4)}$ for a topological system, while in general $I_{(3)} \neq I_{(4)}$ for a system with symmetry-breaking order.

Ultimately, these results depend on and highlight the different global properties of the two systems. Indeed, the bulk of the Kitaev chain is insulating and exponentially decoupled from the boundaries, forcing (by monogamy) the onset of the long-distance correlation between the ends of the chain.
Vice versa, the bulk of the Ising chain is conductive, so that cutting one part of it before performing the partial trace on the reminder abruptly interrupts the information flow and destroys the correlation structure between the two ends of the chain. These physical differences determine the different structure of the end-to-end QCMIs and of the end-to-end SE in the two models. 

Concerning quadripartitions, an interesting, although not surprising, feature is the independence of the results on the relative sizes of the two parts of the bulk $C_2$ (the cut) and $C_1$ (the reminder), so that one can always adopt the most convenient choice $C_1 = 1$ in the actual calculations. 

\begin{figure}
	\includegraphics[scale=0.1]{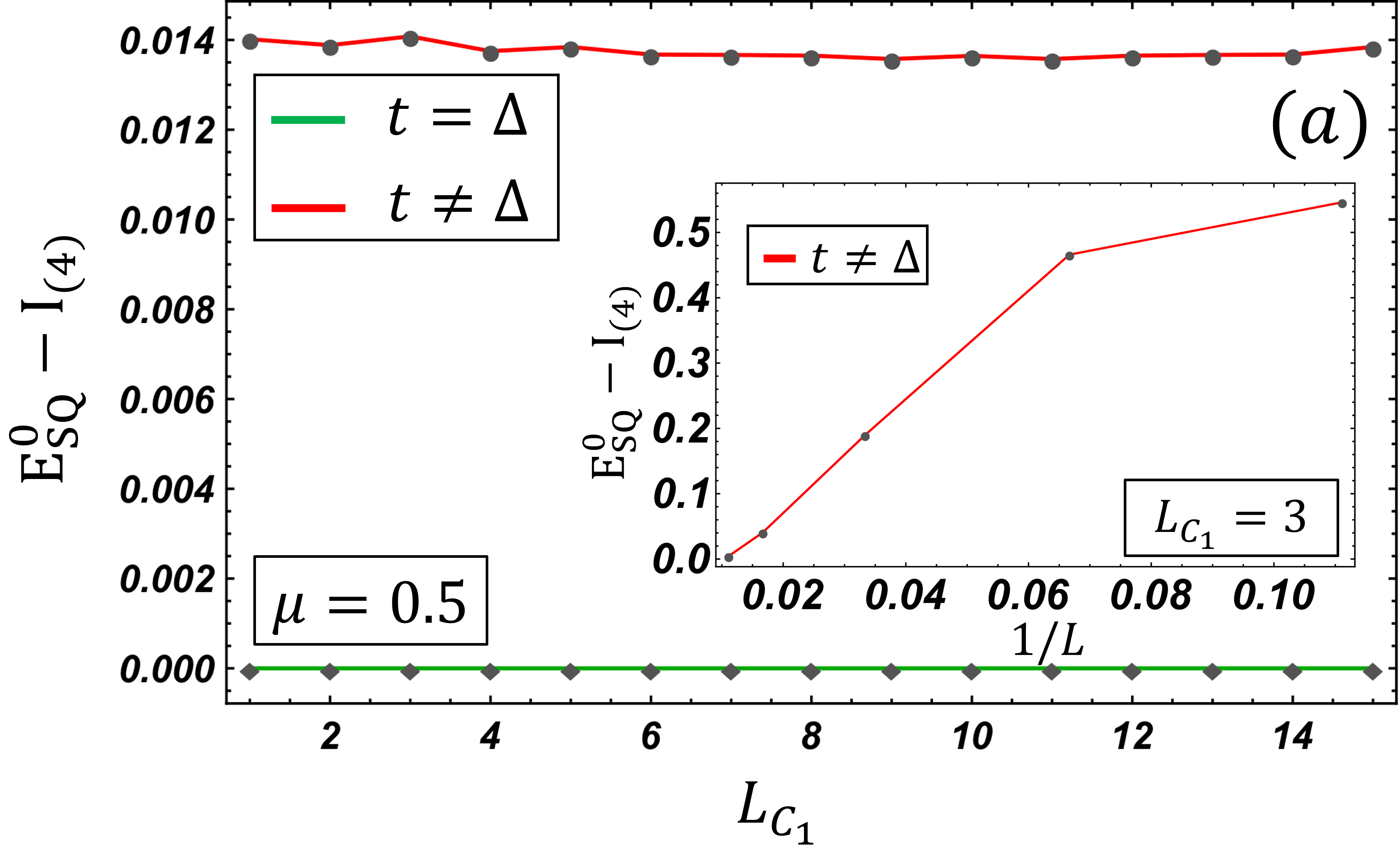}
		\includegraphics[scale=0.2]{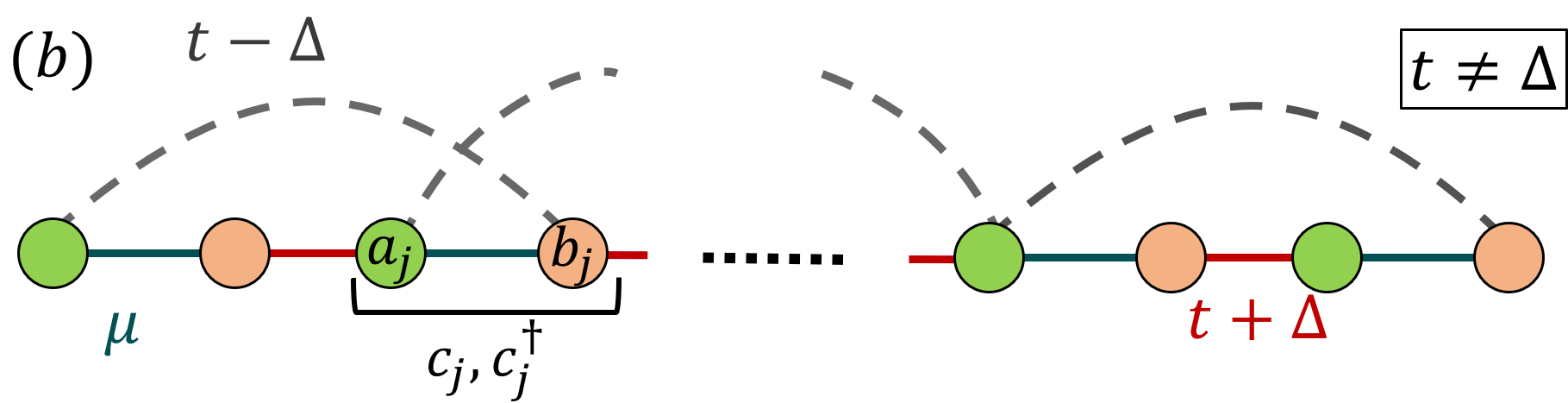}
		\hspace{0.5cm}
			\includegraphics[scale=0.2]{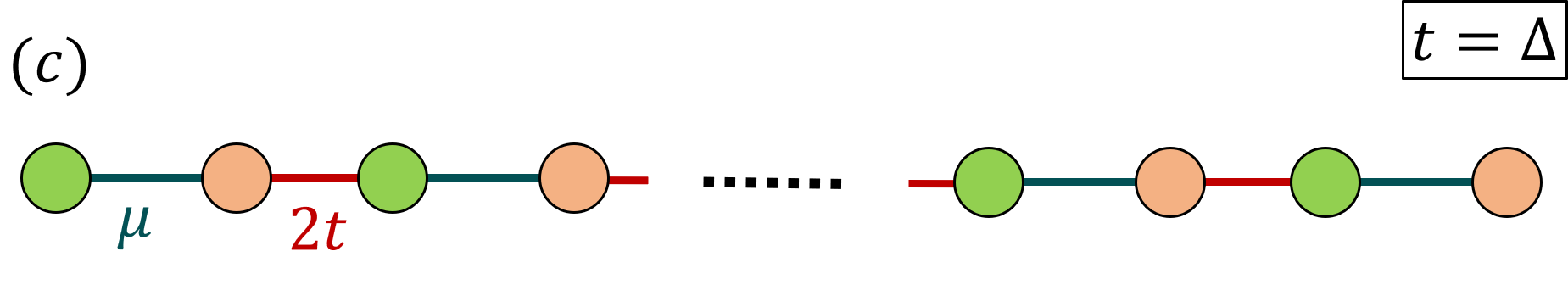}
	\caption{Panel (a): The difference $I_{(4)} - E^0_{sq}$ as a function of the size $L_{C_{1}}$ of the reminder of the bulk in a Kitaev chain of $L=60$ sites. We have considered both the balanced case $t=\Delta=1$ (green curve) and the unbalanced case $t=1$, $\Delta=0.1$ (red curve). Different choices of $L_{C_{1}}$ have no effect on the evaluation of $I_{(4)}$. The inset shows the scaling of $I_{(4)}$ towards the quantized value $E^0_{SQ} = \log(2)/2$ of the TSE in the balanced regime as the size $L$ of the chain is increased. Panel (b): Schematic representation of the chain in the Majorana basis $\{a_j$, $b_j\}$) for $t \neq \Delta$. Panel (c): Same, with $t=\Delta$.}
	\label{Figure5}
\end{figure}

In Fig. \ref{Figure5} panel (a) we report the behavior of the difference $I_{(4)} - E^0_{sq}$ as a function of $L_{C_1}$ with $\mu=0.5$, $t=1$ and $L=60$, and for $\Delta = t$ (green curve) and $\Delta=0.1t$ (red curve). We see that $I_{(4)} - E^0_{sq}$ is constant in both cases, regardless of the length $L_{C_1}$. 

In the balanced hopping-pairing case $t=\Delta$, the SE is fully quantized at the fundamental unit $E^0_{sq} = \log (2)/2$ already for an intermediate chain length of $L=60$ (green curve), while in the strongly unbalanced case $\Delta=0.1t$ one can observe (red curve) a finite-size scaling on $I_{(4)} - E^0_{sq}$. 

These behaviors show that the balanced case $t=\Delta$ is way more insensitive to finite size effects than the unbalanced case $t \neq \Delta$. On the other hand, as the inset in panel (a) of Fig. \ref{Figure5} shows, for the unbalanced case $t \neq \Delta$ the difference $I_{(4)} - E^0_{sq}$ decreases with increasing sizes of the system, tending to vanish, as expected, in the thermodynamic limit. 

The different robustness to finite size effects between the balanced and unbalanced configurations is illustrated in panels (b) and (c) of Fig. \ref{Figure5}, where we draw sketches of the Kitaev chain in the basis of Majorana fermions $a_j=c_j+c^\dagger_j$ and $b_j=-i(c_j-c^\dagger_j)$ for the two aforementioned cases. 
The balanced setting $t=\Delta$ illustrated in panel (c) corresponds to turning off the coupling $(t-\Delta)$ $a_j b_{j+1}$ between third-neighboring Majorana fermions. In this configuration the Majorana fermions are thus more weakly coupled than in the unbalanced case $t \neq \Delta$ illustrated in panel (b). 
On the other hand, the setting $t=\Delta$ and $\mu=0$ corresponds to the exact analytic, topological degeneracy point of the model, where the edge modes $a_1$ and $b_L$ at the two ends of the chain decouple from the lattice and become exact zero energy modes.

\subsection{TSE in the presence of interactions}
\begin{figure}
\includegraphics[width=8.5cm]{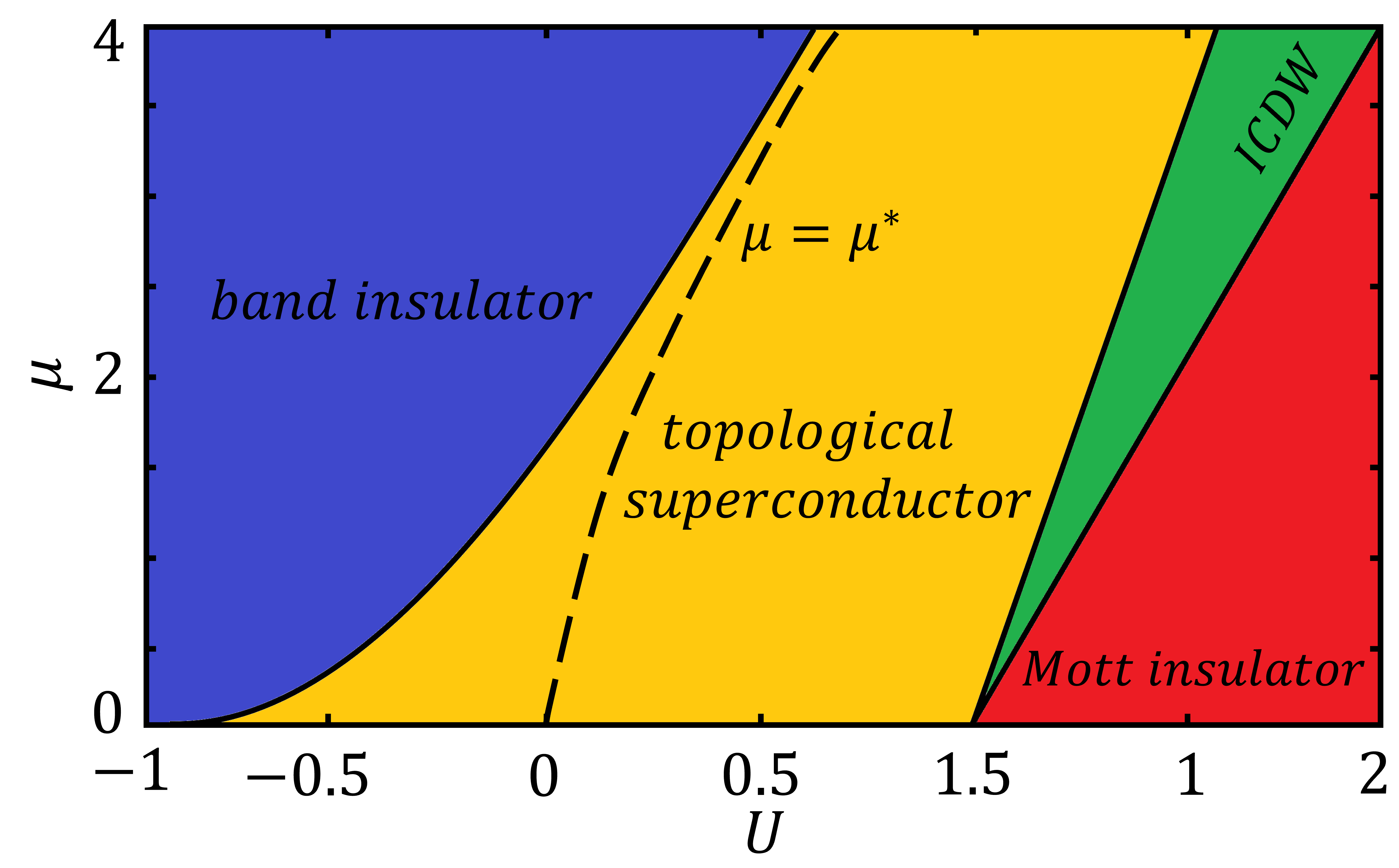}
\caption{Phase diagram of the interacting Kitaev chain on the $\mu$--$U$ plane for $\Delta = t$. The dashed black curve is the factorization line $\mu = \mu^* = 4 \sqrt{U^2+t\ U}$. It starts at the exact topological degeneracy point $\mu = U = 0$, develops entirely inside the topological phase, and pinches asymptotically into the trivial phase for sufficiently large values of $\mu$ and $U$. The acronym $ICDW$ stands for incommensurate charge density wave.}
	\label{Figure6}
\end{figure}

The standard interacting Kitaev chain is obtained by adding a nearest-neighbor density-density interaction term to the free Kitaev Hamiltonian:
\begin{equation}
H_{IK} = H_{K} + U \sum_{j=1}^{L-1} n_j n_{j+1},
\label{interacting}
\end{equation}
where $U$ is the interaction coupling constant and $n_j = c^\dagger_j c_j$ is the on site fermion number operator. The intervals $U > 0$ and $U < 0$ correspond, respectively, to a repulsive and to an attractive interaction. The phase diagram of the model has been investigated in depth by a variety of numerical and approximate analytical methods~\cite{Hassler2012,Sela2011,Dalmonte2020,Katsura2015,PhysRevLett.118.267701} and is reported schematically in Fig.~\ref{Figure6}. 

We see that, although very strong repulsive interactions eventually force the system into a Mott insulating phase, a topological superconducting state is established in between the trivial band insulator phase and the Mott localization even for fairly strong repulsive interactions. The black dashed line inside the topological phase is the factorization line of equation $\mu = \mu^*$ described in the following.

The interacting Kitaev chain can be mapped exactly onto the interacting $XYZ$ spin-1/2 model in an external magnetic field
\begin{eqnarray}
H_{xyz} & = & \sum_{j=1}^{L-1} \left[ J_x \sigma_j^x \sigma_{j+1}^x + J_y \sigma_j^y \sigma_{j+1}^y + J_z \sigma_j^z \sigma_{j+1}^z \right] \nonumber \\
& + & h\sum_{j=1}^L \sigma_j^z \, , 
\label{XYZ}
\end{eqnarray}
where $J_x = - (\Delta + t)/2$, $J_y = (\Delta - t)/2$, $J_z = - U$, and $h = \mu/2$. 

The model admits exact ground state solutions factorized in the product of single-spin (single-site) wave functions along the so-called factorization line $h = h^* = z_d \sqrt{(J_x + J_z) (J_y + J_z)}$ in any dimension~\cite{Illuminati2008,Illuminati2009,Montangero2013}, with $z_d$ the coordination number (so that, e.g., $z_d = 2$ in the 1D case).

For the Kitaev chain, this corresponds to the exact solution for the many-body ground states along the factorization line~\cite{Katsura2015}:
\begin{equation}
\mu=\mu^* = 4 \sqrt{U^2+t\ U+(t^2-\Delta^2)/4} \, ,
\label{factorizationline}
\end{equation}
which reduces to $\mu^* = 4\sqrt{U(U + t)}$ for $t = \Delta$.

In general, along any factorization line (provided it exists) an interacting model Hamiltonian $H$ becomes frustration--free and can be written as the sum of commuting local terms $H_j$: $H = \sum_j H_j$ with $[H_i,H_j]=0$~\cite{Illuminati2010}. 

The exact ground states of definite symmetry of the interacting Kitaev chain for $\mu = \mu^*$ are the following entangled linear combinations~\cite{Katsura2015}: 
\begin{eqnarray}
&&|\Psi^{even}\rangle =\frac{1}{\sqrt{2}} \bigl(|\Psi^{+}\rangle + |\Psi^{-}\rangle \bigr) \\
&&|\Psi^{odd}\rangle  = \frac{1}{\sqrt{2}} \bigl(|\Psi^{+}\rangle - |\Psi^{-}\rangle \bigr),
\label{exact}
\end{eqnarray}  
where the non-orthogonal states $|\Psi^{+}\rangle$ and $|\Psi^{-}\rangle$ are fully factorized in the product of single-site wave functions and read~\cite{Katsura2015}:
\begin{equation}
|\Psi^{\pm}\rangle = \frac{1}{(\alpha+1)^{L/2}}e^{\pm \alpha c^{\dagger}_1} \dots\ e^{\pm \alpha c^{\dagger}_L} |vac\rangle ,
\label{factorized}
\end{equation}
where
\begin{equation}
\alpha=\sqrt{\cot(\theta^*/2)},\ \ \theta^*=\arctan(2 \Delta/\mu^*) ,
\label{alfa}
\end{equation}
with $|vac\rangle$ denoting the lattice vacuum state.

The ground states $|\Psi^{even}\rangle$ and $|\Psi^{odd}\rangle$ are orthogonal and are the only possible ground state of $H_{IK}$ for $\mu=\mu^*$. Ground state factorization holds in the topological phase with a repulsive interaction $U \geq 0$, and the even and odd ground states are degenerate with energy $E_0=-(L-1)(U+t)$. 

\begin{figure}
	\includegraphics[scale=0.12]{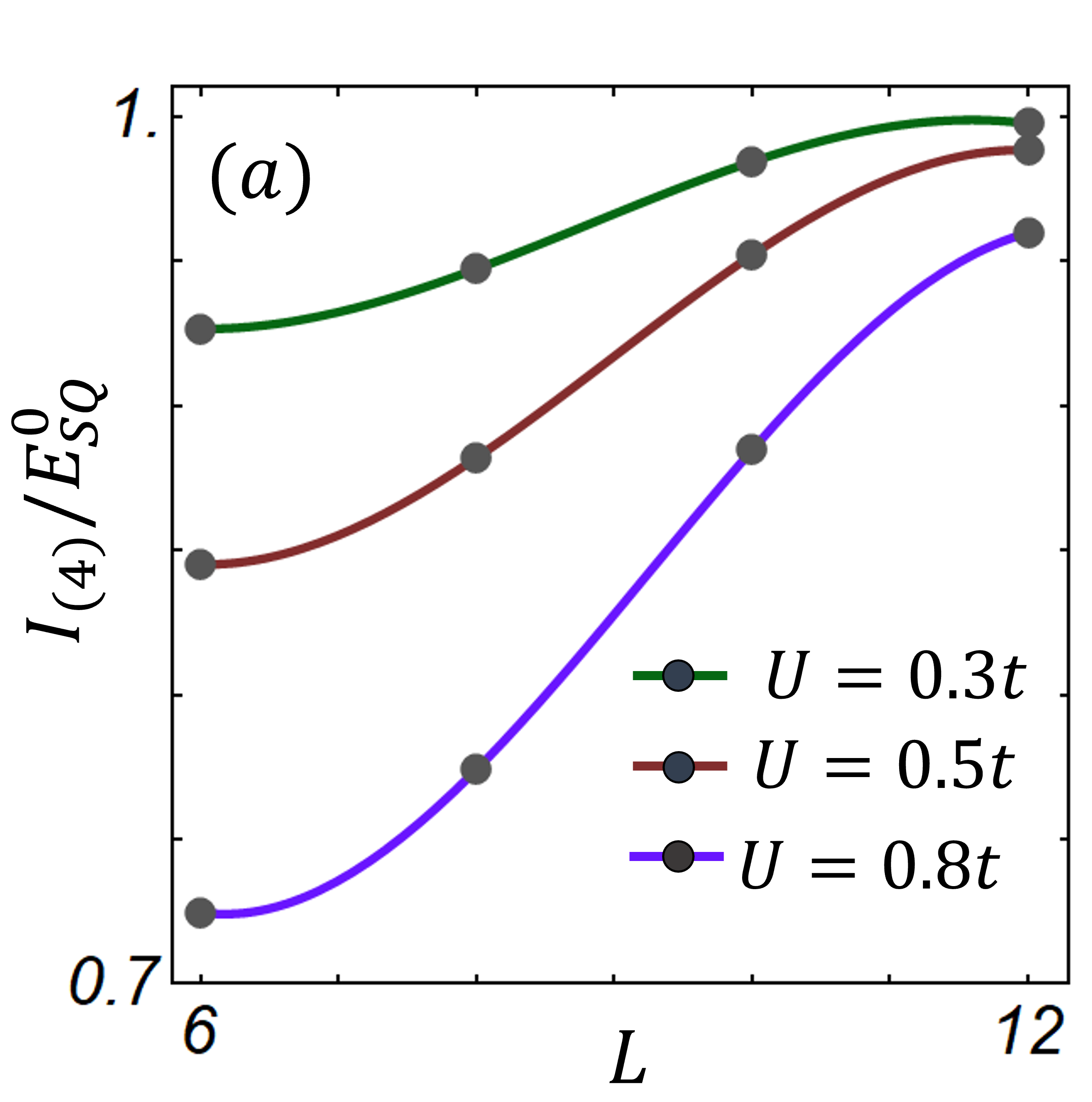}
			\includegraphics[scale=0.12]{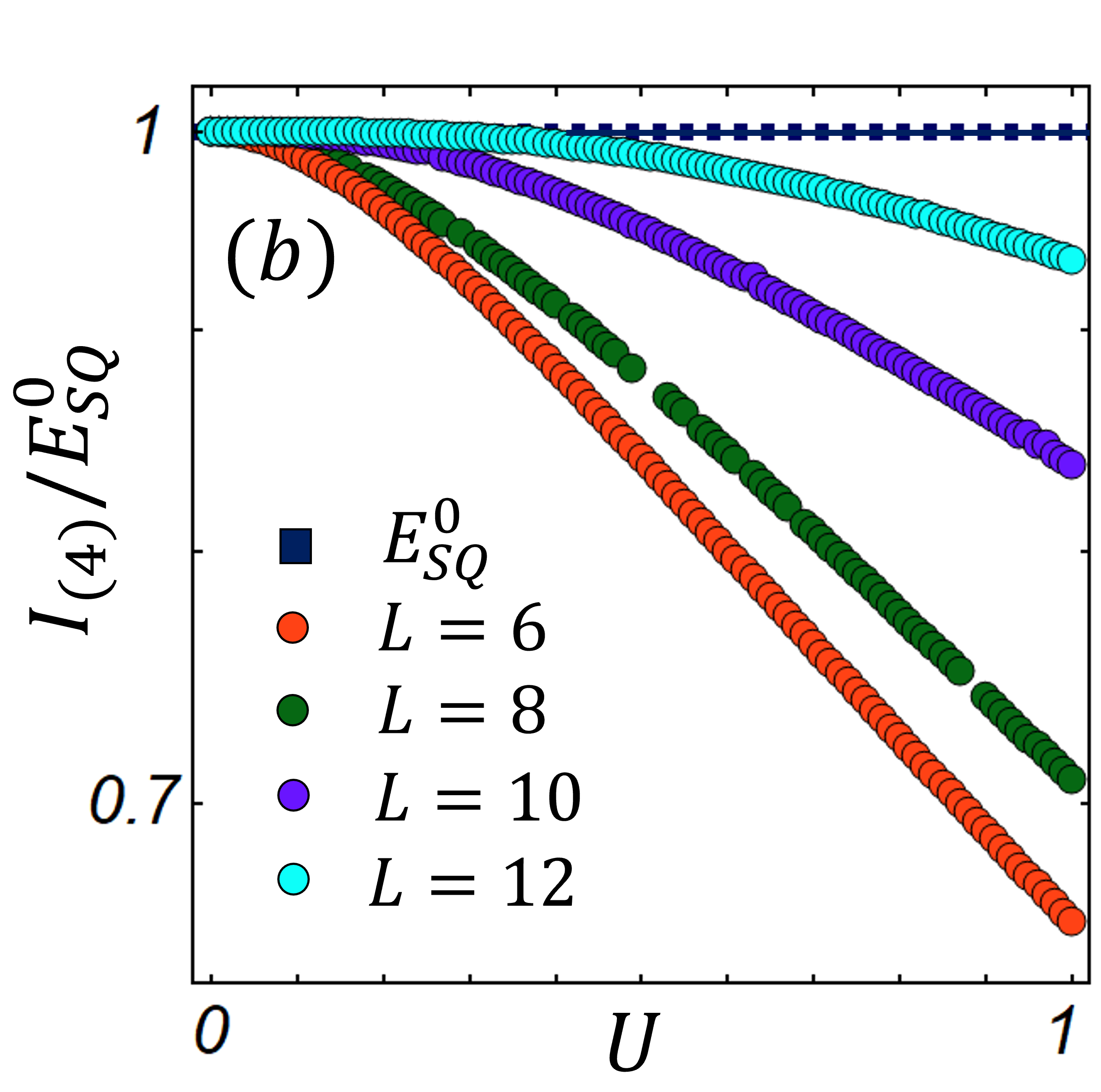}
	\caption{The QCMI to TSE ratio $I_{(4)}/E^0_{sq}$ for the exact ground state solutions of the interacting Kitaev chain on the Illuminati-Katsura factorization line $\mu = \mu^*$ as a function of the chain length $L$ for different values of the interaction strength $U$ (panel (a)), and as a function of $U$ for different values of $L$ (panel (b)). In both cases the scaling behavior of $I_{(4)}$ converges to the quantized value $log(2)/2$ of the TSE for the non-interacting Kitaev chain. Eventually, for increasingly large values of $U$, the factorization line pinches into the trivial phase and the QCMI $I_{(4)}$ vanishes asymptotically. Throughout, $t=\Delta$ and $L_e=L/3$.}
	\label{Figure7}
\end{figure}

The behavior of the QCMI $I_{(4)}$ normalized to the quantized unit of TSE $E^0_{sq} = log(2)/2$ along the Illuminati-Katsura factorization line $\mu = \mu^*$ is reported in Fig.~\ref{Figure7} panel (a) as a function of the chain length $L$ for different values of the interaction strength, and in panel (b) as a function of the interaction strength for different values of the chain length $L$. 

The QCMI $I_{(4)}$ exhibits the correct scaling, saturating to the quantized value $log(2)/2$ of the SE for sufficiently large values of the chain size $L$, and vanishing asymptotically for very large values of the interaction strength $U$ as the factorization line pinches the boundary of the topological phase and the system enters in the trivial phase. 

In conclusion, we see that the quantized TSE $E^0_{sq}$ identifies the correct nonlinear order parameter for the topological phase of the fermionic Kitaev chain, either free or interacting.

\subsection{Robustness of TSE against disorder}
Disorder plays a fundamental role in low-dimensional electronic systems \cite{Brouwer2011,PhysRevMaterials.5.014204,PhysRevB.104.195117,Hegde2016} and robustness against disorder is a defining property of topological materials. In fact, disorder may even induce localization effects as in the case of Anderson insulators \cite{PhysRevLett.102.136806}, thus favoring topological phases of matter, whereas in other systems such as, e.g., semiconductor Majorana nanowires and topological insulator nanoribbons, it can yield detrimental effects \cite{PhysRevMaterials.5.014204,PhysRevMaterials.5.124602}. It has been shown that topological phases of the Kitaev chain are robust to the effects of disorder and local perturbations \cite{Brouwer2011,Neven2013,Adagideli2014,Hegde2016}. Such immunity is a fundamental property that is expected to hold whenever topological superconductors are involved. This observation suggests that the edge-to-edge QCMIs and SE should also be as robust to disorder as other topological indicators defined in terms of spectral or transport properties \cite{Sticlet2012,Beenakker1997}.

We study two distinct classes of configurations, each characterized by a different source of disorder with a clear tracking of its physical origin. Since one of the prominent schemes to realize a topological Kitaev chain is by proximity effects between a semiconducting nanowire and a conventional superconductor \cite{Lutchyn2010,Oreg2010}, we consider two distinct sources of disorder: random site-dependent hopping amplitudes $t_i$ and 
random site-dependent pairing potentials $\Delta_i$. The former provides a model of the effective mass gradient and random doping along the nanowire that originate from the growth process of the nanowire itself; the latter emulates unwanted spatial variations along the wire which can affect the nanowire-superconducting coupling and therefore the induced superconducting gap.

\begin{figure}
	\includegraphics[scale=0.09]{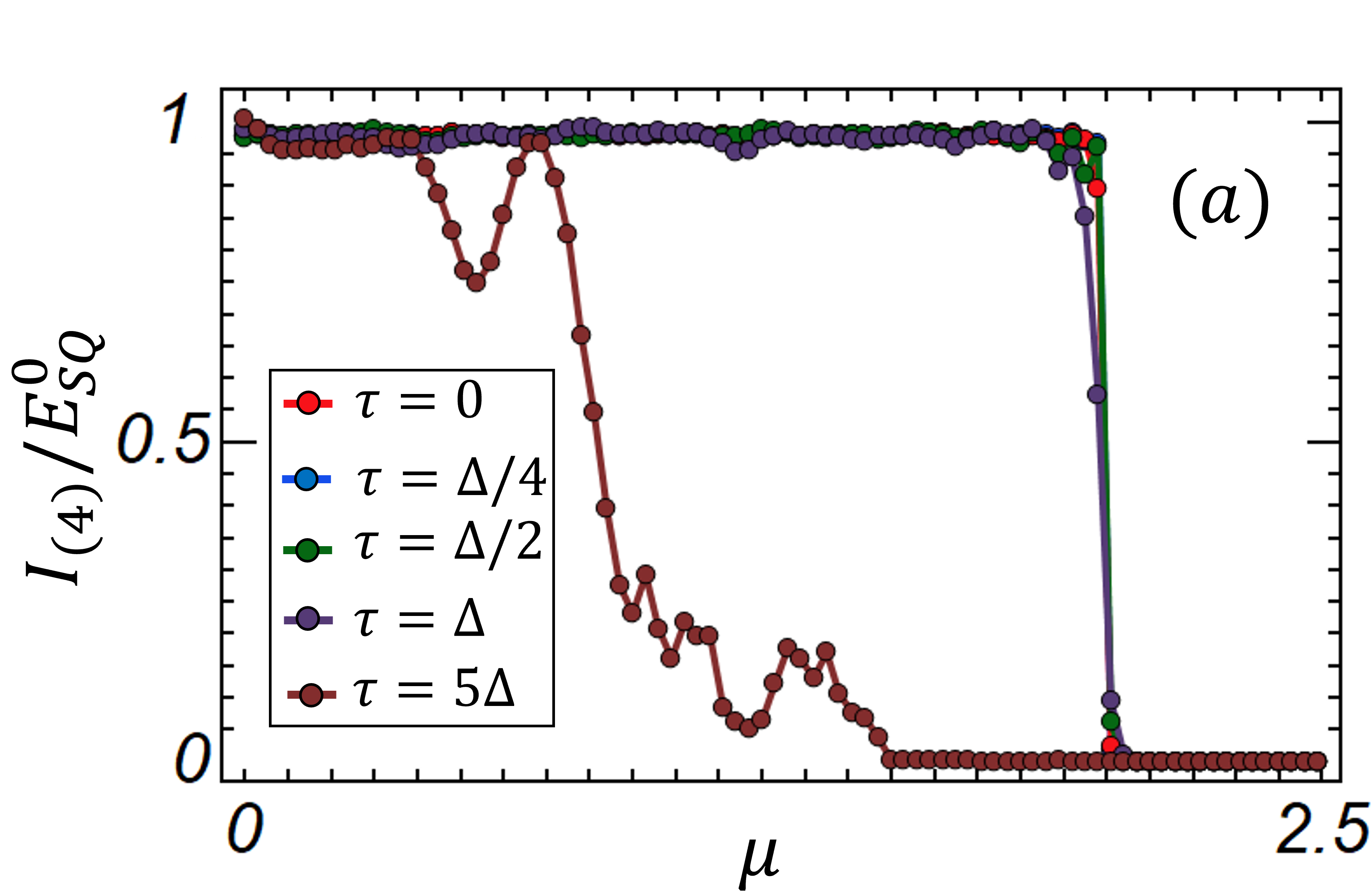}\\
	\vspace{0.3cm}
	\includegraphics[scale=0.09]{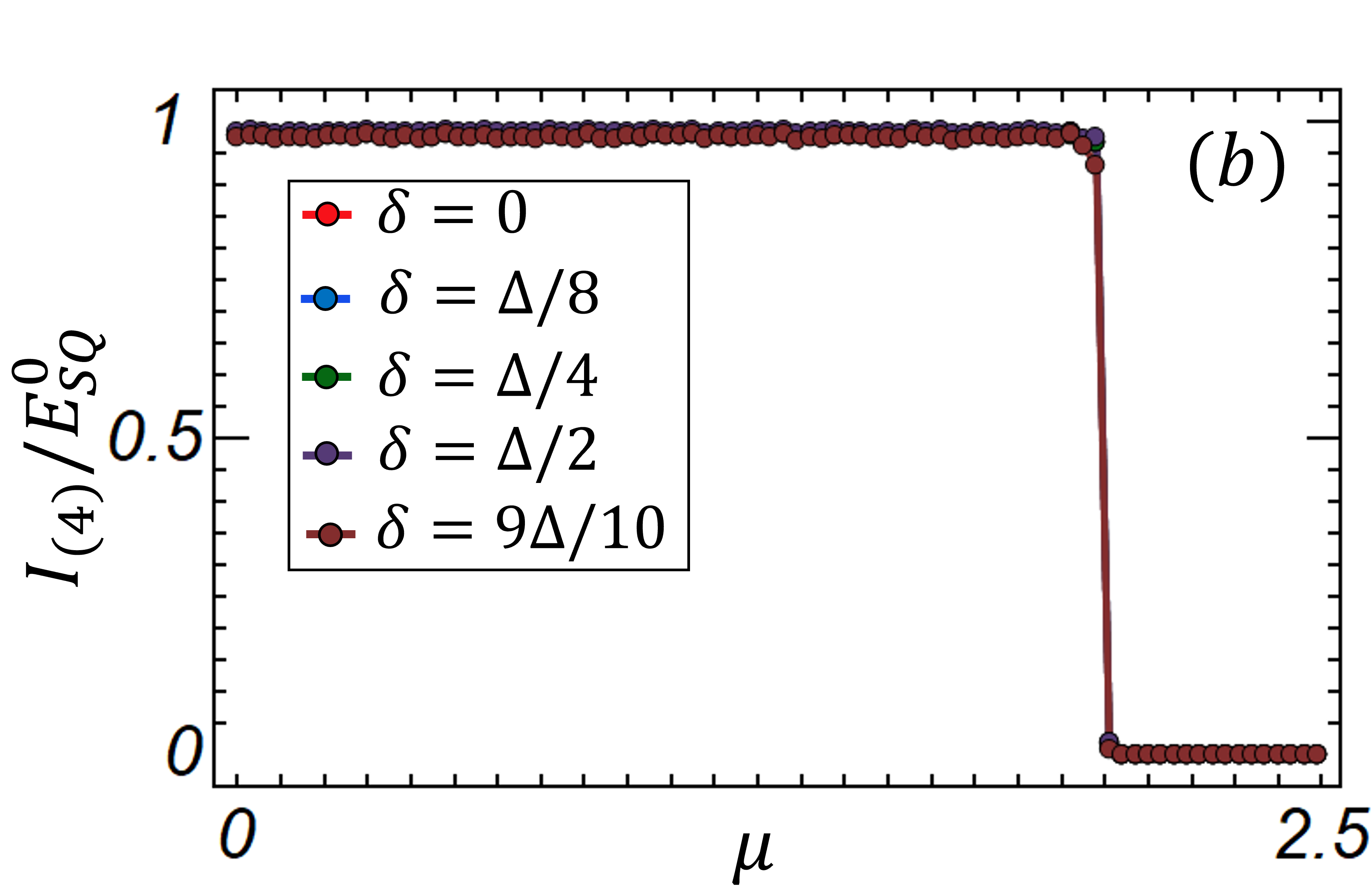}\\
			\caption{The QCMI to TSE ratio $I_{(4)}/E^0_{sq}$ as a function of the chemical potential $\mu$ for a chain of length $L = 70$ and two different types of disorder. Panel (a): behavior of $I_{(4)}/E^0_{sq}$ in the presence of random hopping amplitudes $t_i$. Panel (b): behavior of $I_{(4)}/E^0_{sq}$ in the presence of random pairing potentials $\Delta_i$. The random realizations are generated by a continuous probability distribution defined in the interval $t_i \in (t-\tau$, $t+\tau)$, $\Delta_i \in (\Delta-\delta$, $\Delta+\delta)$. The reference values of the parameters have been fixed as $t=1$, $\Delta=0.1$. The different curves are parameterized by disorder of increasing strength.}
			\label{Figure8}
		\end{figure}

In Fig.~\ref{Figure8} we report the behavior of the QCMI to TSE ratio $I_{(4)}/E^0_{sq}$ of the Kitaev chain Eq.~\ref{Kitaev} as a function of the chemical potential $\mu$ for the aforementioned two different sources of disorder. In panels (a) and (b) we consider respectively random hopping integrals $t_i=t+\tau^{dis}_i$ and random pairing potentials $\Delta_i=\Delta+\delta_i^{dis}$ with the random strengths of disorder $\tau^{dis}_i$ and $\delta_i^{dis}$ uniformly distributed in the intervals $(-\tau$, $\tau)$ and $(-\delta$, $\delta)$.

In Fig.~\ref{Figure8}, for the sake of completeness disorder effects are investigated spanning the spectrum of noise strengths from the perturbative regime up to strongly disordered configurations. It is worth mentioning here that the maximum disorder strength considered in panel (b) of Fig. \ref{Figure8} is comparable with the mean value of the pairing potential $\Delta$. Such configuration corresponds to an extremely disordered system, a situation that is not expected in realistic experimental conditions. A similar phenomenology can be observed by studying disorder effects induced by random values of the hopping integral, see Fig. \ref{Figure8} panel (a), from $\tau=\Delta/4$ up to $\tau=\Delta$ while for $\tau=5 \Delta$ the phase boundary defining the topological transition is reduced. This behaviour suggests that a random hopping is more effective in perturbing topologically ordered phases than a random superconducting pairing. 

The actual values of disorder experimentally achievable with currently available technologies are in general significantly smaller than those simulated in Fig. \ref{Figure8}; the latter must thus be taken as a theoretical-numerical test/prediction of the robustness of edge QCMI and edge TSE. In particular, the value $\tau=5\Delta$, although not comparable with the maximally allowed disorder on the energy gap, is of the order of half of the band width; therefore, it must certainly be considered nonperturbative.

Indeed, Fig.~\ref{Figure8} shows, as it should be expected, that within the typical regimes of weak to moderate disorder the TSE nonlocal order parameter is strongly resilient in the entire superconducting phase.  

\section{Systems with multiple topological phases and edge modes: TSE of the Kitaev ladder} 
\label{KL}

In this section we generalize the previous investigations to consider systems enjoying multiple topological phases and multiple MZEMs, i.e. more than one Majorana zero mode per edge.

Various such generalizations of the $1D$ Kitaev model have been introduced \cite{Wimmer2010,Potter2010,Zhou2011,Wakatsuki2014,Schrade2017,Zhou2017}; the simplest instance can be realized by coupling multiple Kitaev chains with transverse hopping and pairing terms to form quasi one-dimensional $n$-leg Kitaev ladders~\cite{Maiellaro2018,Maiellaro2019}. The case $n=2$ defines the two-leg Kitaev ladder. This is the minimal model featuring multimode topological phases, which usually arise in models of significant complexity that involve higher-dimensional platforms \cite{Benalcazar2017,PhysRevA.101.063839,Maiellaro2021bis,Maiellaro2022}. 

The Hamiltonian $H_{KL}$ of the two-leg Kitaev ladder can be written as follows:
\begin{equation}
H_{KL} = H_{K_1} + H_{K_2} + H_{K_{12}} \, ,
\label{ladder}
\end{equation}
where the two Kitaev chains $H_{K_1}$ and $H_{K_2}$ define the legs of the ladder and 
\begin{equation}
H_{K_{12}} = \sum_{i=1}^{L} [-t_1 c^\dagger_{i,1}c_{i,2} + \Delta_1 c_{i,1}c_{i,2} + h.c.] 
\label{legs}
\end{equation}
describes the coupling between the two legs $H_{K_1}$ and $H_{K_2}$. The coupling is realized by means of a transverse hopping with amplitude $t_1$ and a transverse pairing with amplitude $\Delta_1$; these two transverse Hamiltonian terms provide the rungs of the ladder. 

The model has been introduced and discussed at length in Ref.~\cite{Maiellaro2018}. The Kitaev ladder satisfies particle-hole, time reversal and chiral symmetry, belonging to the $BDI$ class of the ten-fold classification. 
\begin{figure}
	\includegraphics[scale=0.1]{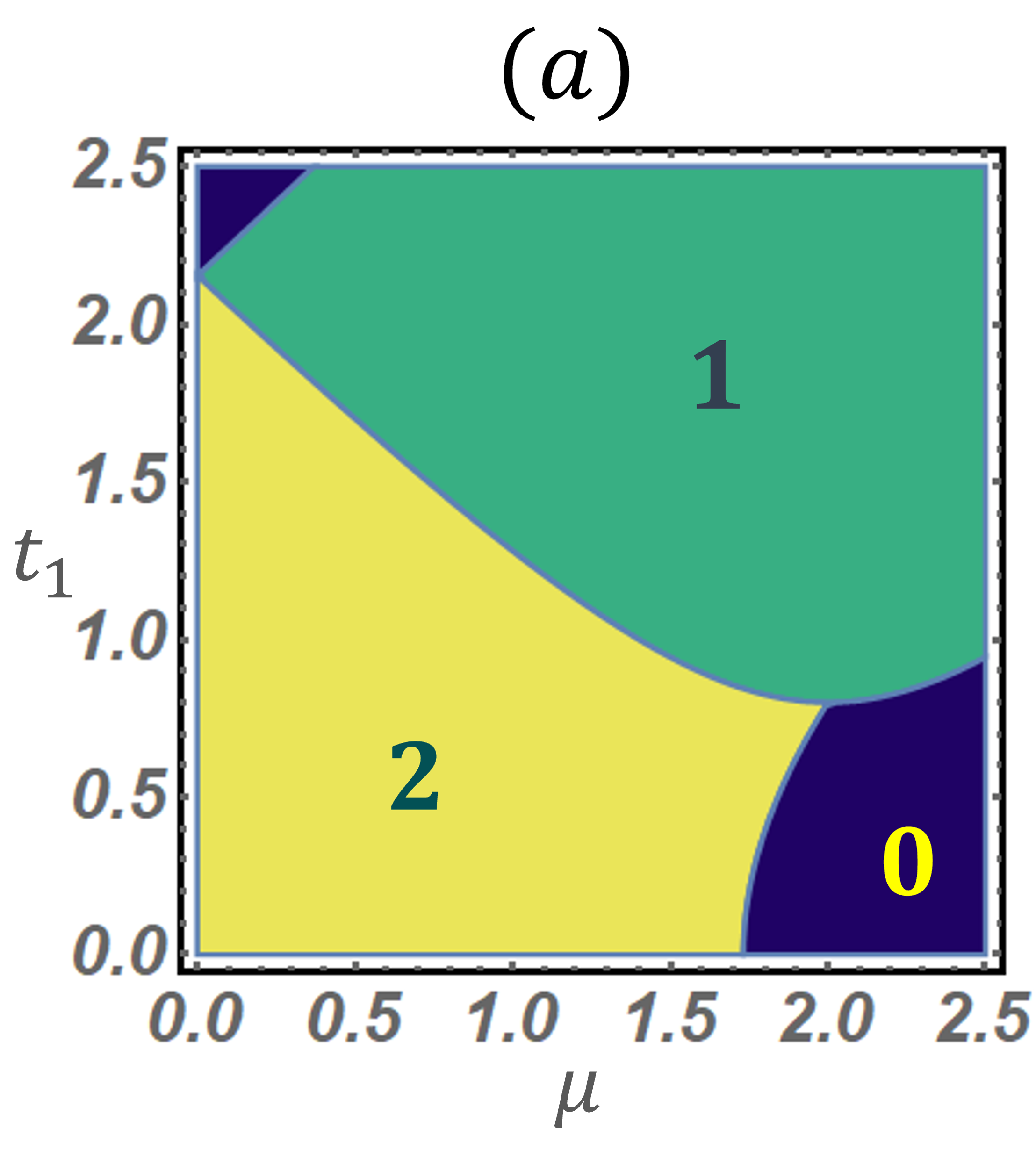}
	\hspace{0.1cm}
	\includegraphics[scale=0.1]{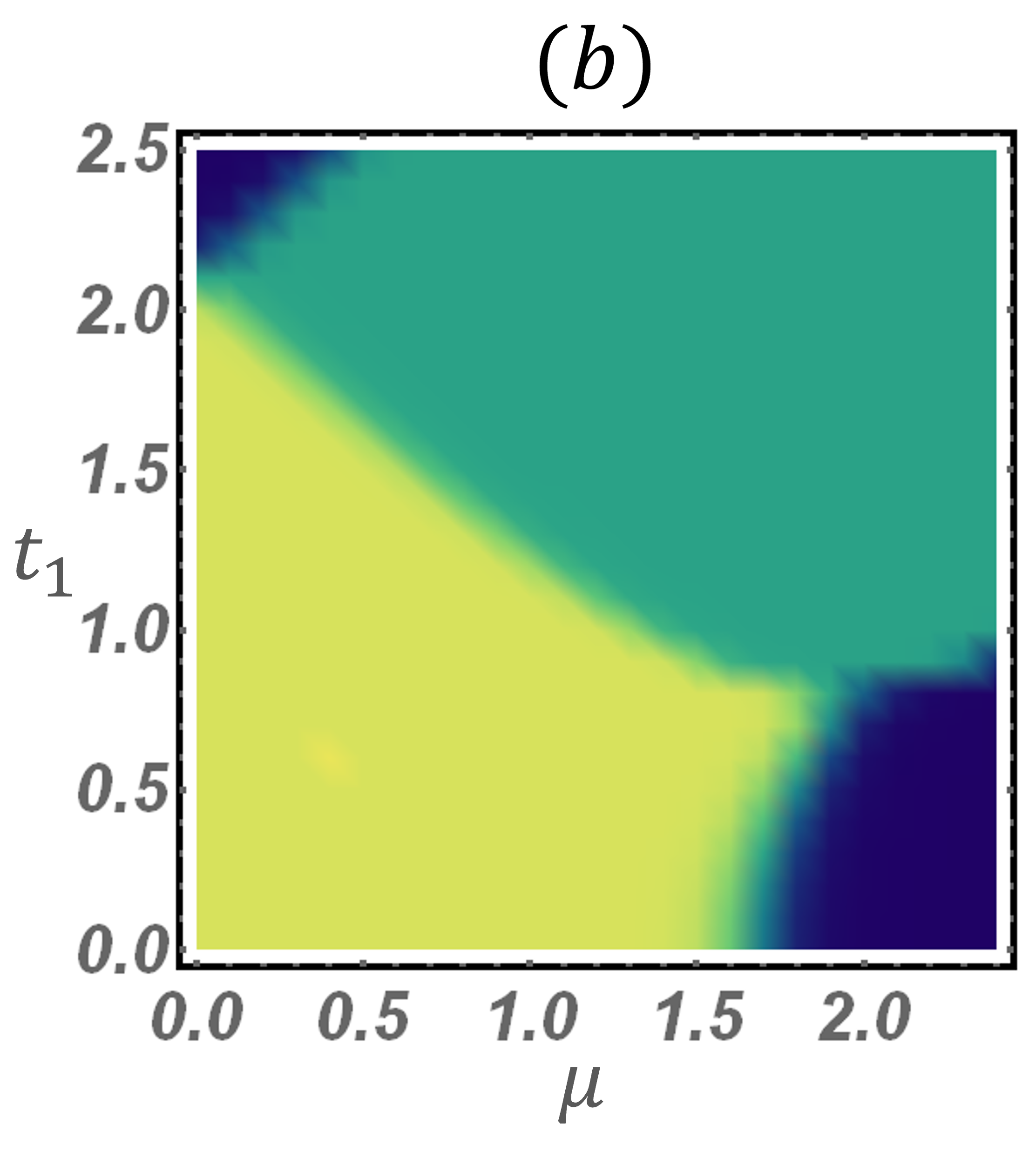}\\
	\vspace{0.5cm}
	\includegraphics[scale=0.1]{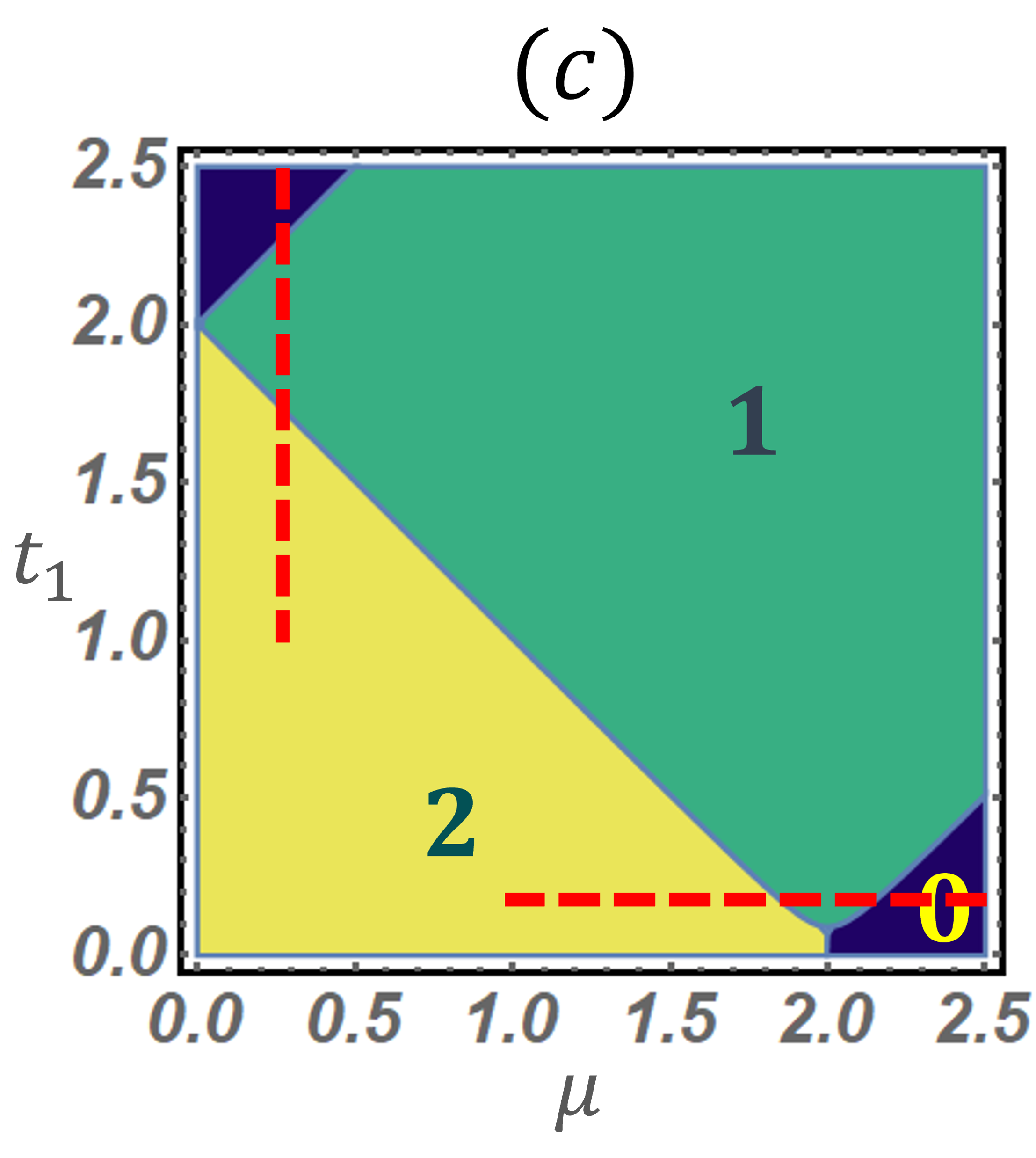}
	\caption{Panels (a) and (c): phase diagram of the two-leg Kitaev ladder determined by means of the winding number in $k$-space. Panel (a): phase diagram in the $\mu$--$t_1$ plane, with the remaining Hamiltonian parameters fixed at $t=1$, $\Delta=0.8$, and $\Delta_1=0.8$. Panel (c): phase diagram in the $\mu$--$t_1$ plane, with the remaining Hamiltonian parameters fixed at $t=1$, $\Delta=0.8$, and $\Delta_1=0.09$. Panel (b): phase diagram of the two-leg Kitaev ladder determined by means of the QCMI to TSE ratio $I_{(4)}/E^0_{sq}$. The phase diagram is in the same $\mu$--$t_1$ plane, with the values of the remaining Hamiltonian parameters set as in panel (a), and it is obtained on a grid of $25 \times 25$ points with the order of the interpolating polynomial equal to $1$. The dashed red lines in panel (c) denote the straight transition lines ("cuts") that cross the different phase boundaries. The horizontal cut is at constant transverse hopping $t_1$; the vertical cut is at constant chemical potential $\mu$. In all panels, the length $L$ per ladder leg and the length $L_e$ per edge partition are fixed, respectively, at $L=50$ and $L_e=12$.}
\label{Figure9}
\end{figure}

The phase diagram of the model can be investigated by means of the winding number:
$$W=Tr\int^\pi_{-\pi} \frac{dk}{2\pi i}\ \ A_k^{-1} \partial_k A_k=$$
\begin{equation}
	-\int^\pi_{-\pi}\frac{dk}{2\pi i}\ \ \partial_k\ln Det A_k
\end{equation}\\
where $A_k$ can be expressed in terms of the model parameters once we impose periodic boundary conditions and the bulk-edge correspondence is invoked:
\begin{eqnarray}
	A_k=
	\begin{pmatrix}
		\epsilon_k-\Delta_k&-t_1-\Delta_1\\
		-t_1+\Delta_1&\epsilon_k-\Delta_k\\
	\end{pmatrix},
\end{eqnarray}
where $\epsilon_k=2t\cos k+\mu$, $\Delta_k=2i\Delta \sin k$ and $k \in [-\pi,\pi]$.

In Fig.~\ref{Figure9} we report the phase diagrams of the two-leg Kitaev ladder obtained respectively by means of the winding number, as shown in panels (a) and (c), and by means of the QCMI $I_{(4)}$ normalized by the quantized edge-edge TSE $E^0_{sq}$, as shown in panel (b) for the same choice of Hamiltonian parameters as in panel (a). The diagram in panel (b) is obtained for a ladder of $L=50$ sites per leg (chain), with the length of each edge partition fixed at $L_e=12$, and via an interpolation on a grid of $25 \times 25$ points. 

The digits drawn on the phase diagrams in panels (a) and (c) count the number of MZEMs per ladder edge in each phase. Indeed, the system possesses a richer phase diagram compared to that of the single Kitaev chain. Denoting by $N_{m}$ the number of MZEMs per edge, the two-leg Kitaev ladder features two alternating topological phases, respectively endowed with $N_{m} = 1$ and $N_{m} = 2$ edge modes, and a trivial phase with no edge modes ($N_m = 0$). 

Despite the fact that the phase diagrams in panels (a) and (c) are defined in terms of bulk properties, while the phase diagram in panel (b) is obtained in terms of the topological boundary entanglement, they show an excellent qualitative and quantitative agreement already at moderate system sizes. 

In panel (c), we report the contour plot of the winding number for a different set of values of the Hamiltonian parameters with respect to the ones of panels (a) and (b). Indeed, with the choice of Hamiltonian parameters as in panel (c), the system features the possibility of moving along straight transition lines that cross all the three different phases $2$-$1$-$0$, once either $t_1$ or $\mu$ is kept constant. The dashed red lines drawn in panel (c) denote such horizontal (constant $t_1$) and vertical (constant $\mu$) cuts across the different phase boundaries.

\begin{figure}
    \includegraphics[scale=0.1]{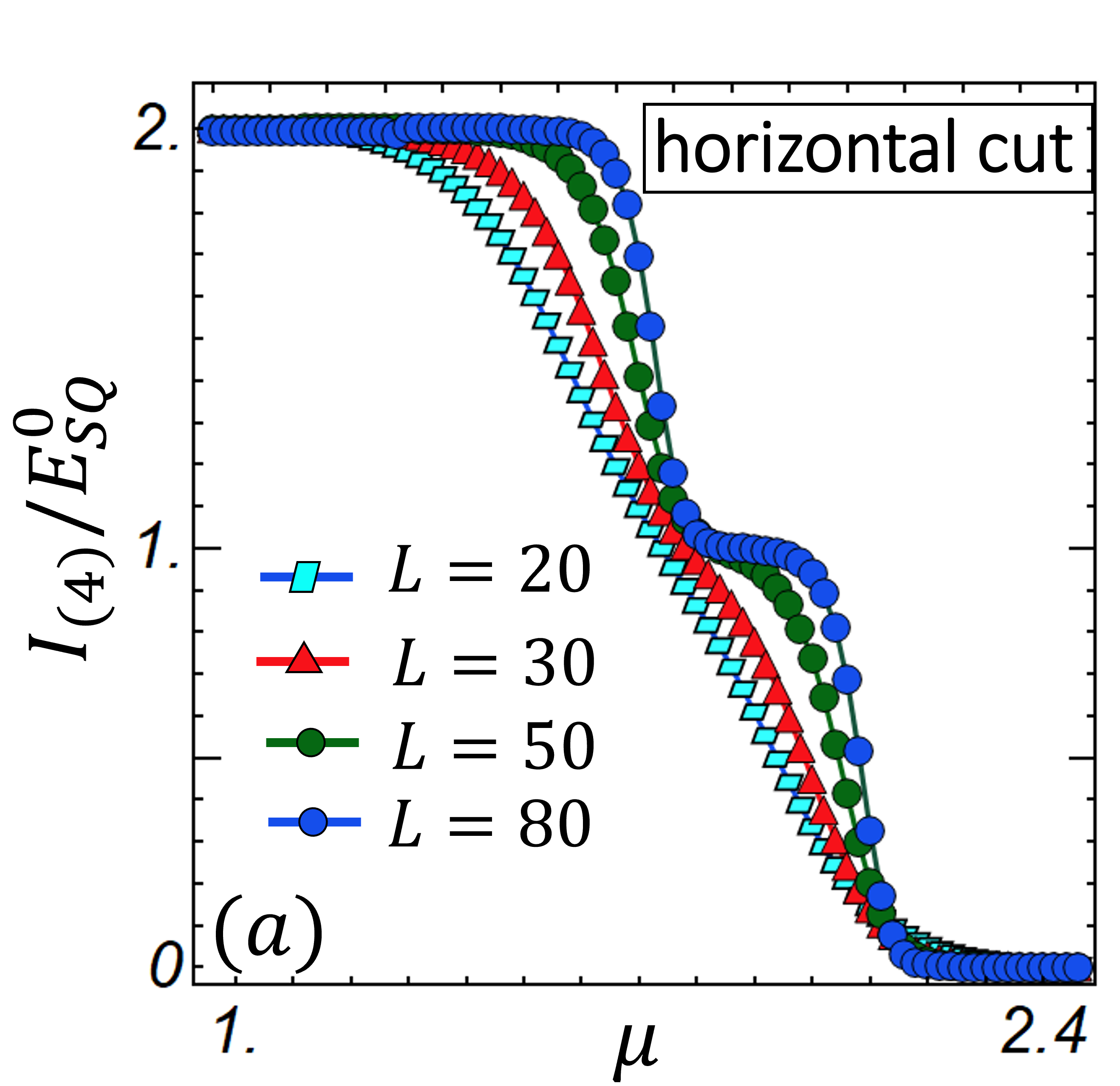} \\
		\vspace{0.5cm}
	\includegraphics[scale=0.1]{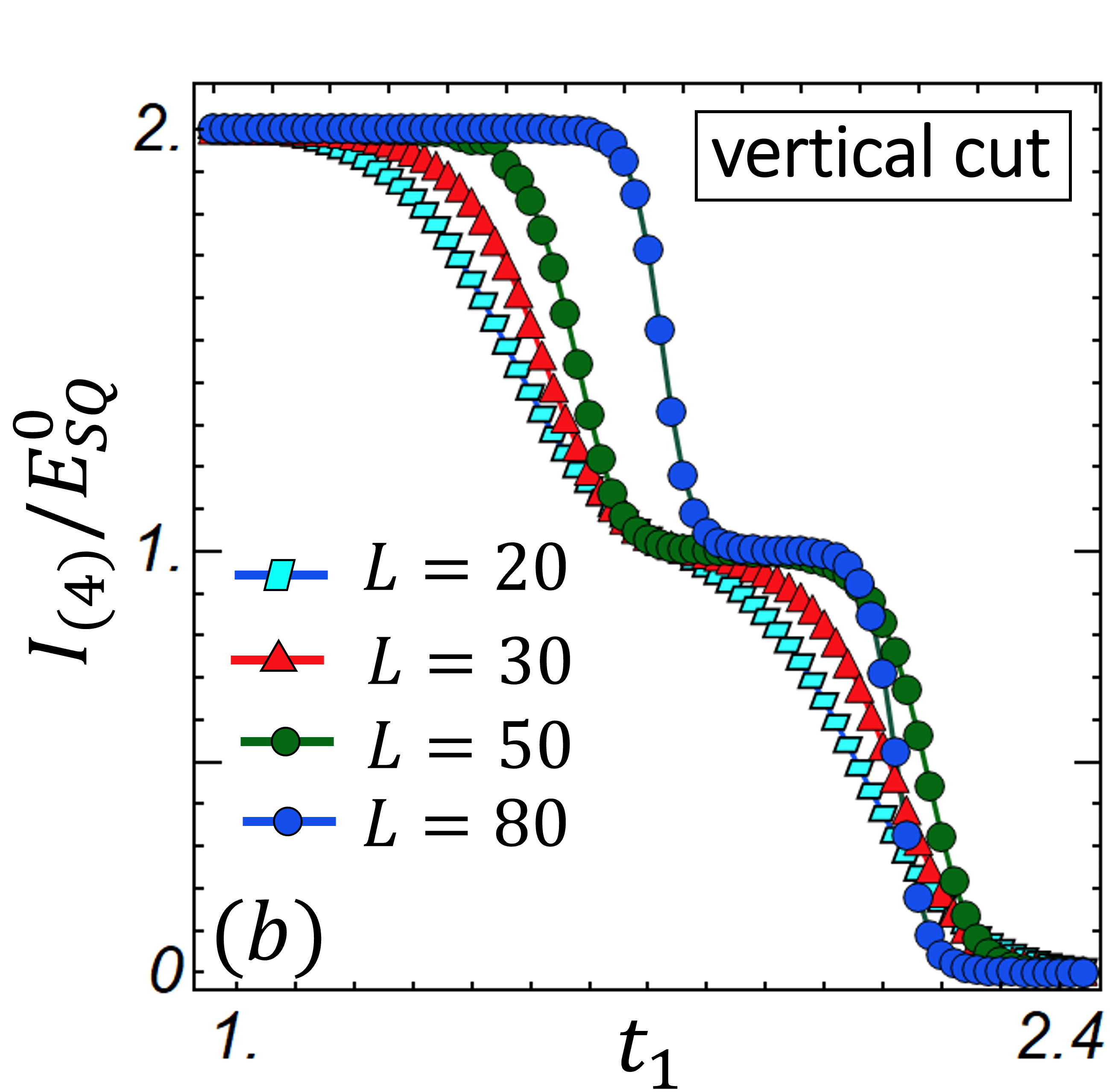}
	\caption{Panel (a): QCMI-TSE ratio $I_{(4)}/E^0_{sq}$ as a function of the chemical potential $\mu$ along the horizontal cut at fixed transverse hopping $t_1$ drawn in panel (c) of Fig.~\ref{Figure9}, for different lengths $L$ of the ladder. Panel (b): $I_{(4)}/E^0_{sq}$ as a function of $t_1$ along the vertical cut at constant $\mu$ drawn in panel (c) of Fig.~\ref{Figure9}, for different values of the length $L$ of any of the two equal legs of the ladder. The plots at different lengths of the ladder legs show the fast scaling behavior of the QCMI-TSE ratio with the system size. Irrespective of the choice of the transition line crossing the different phase boundaries, in each phase $I_{(4)}/E^0_{sq}$ is quantized to an integer counting the number of edge MZEMs featured by the phase, and jumps by one unit at each transition, realizing a series of Hall-like plateaux.}
\label{Figure10}
\end{figure}

In Fig.~\ref{Figure10} we report the behavior of the QCMI-TSE ratio $I_{(4)}/E^0_{sq}$, respectively along the horizontal and vertical cuts crossing the different phases of the ladder, as drawn in  panel (c) of Fig.~\ref{Figure9}. In both cases, we observe the three plateaux $I_{(4)}/E^0_{sq} = 2, 1, 0$ corresponding, respectively, to the two topological phases and to the trivial phase. 

Remarkably, we find that TSE not only discriminates topological phases from trivial ones but also distinguishes different topological phases by counting the corresponding number of Majorana edge modes $N_{m}$. Indeed, for the two-leg Kitaev ladder we have:
\begin{equation}
\frac{I_{(4)}}{E^0_{sq}} = N_{m} \, .
\label{counting}
\end{equation}

If the above relation has general validity, then in principle QCMI and TSE could apply to topological systems of arbitrary complexity featuring any number of topological modes. Besides its conceptual importance, this property bears a clear practical advantage, since while the definition of the TSE is unique and system-independent, different invariants have to be defined each time, following the ten-fold classification, depending on the symmetries obeyed by each specific system under investigation.

Another paradigmatic model belonging to the same BDI topological class of the two-leg Kitaev ladder is the Su-Schrieffer-Heeger (SSH) fermionic chain \cite{PhysRevLett.42.1698}. For such system the disconnected R\'enyi entropy has been shown to reproduce correctly the phase diagram of the model \cite{10.21468/SciPostPhysCore.3.2.012}.

Despite belonging to the same topological universality class and therefore sharing the same topological band invariant (i.e., the winding number), the two models feature different phase diagrams and, perhaps more importantly, differ in the statistics of the edge modes. In particular, while the SSH model exhibits a trivial phase and one topological phase with one fermionic mode per edge, the two-leg Kitaev ladder features a trivial phase and two different topological phases, respectively with one and two Majorana modes per edge. Therefore, in one case (SSH) we have a simple phase diagram with one topological phase and fermionic edge statistics, while in the other case (ladder) we have a richer phase diagram with two different topological phases and nonabelian anyon edge statistics.
 
What motivates the investigation of the relatively simple example of the two-leg superconducting ladder is precisely that it allows to study the transition between different multi-mode topological phases and to verify that our scheme indeed discriminates between them by the occurrence of different quantized plateaux of the TSE that count the integer number of edge modes present in each phase. These results indicate that our scheme is in principle also applicable to any multi-mode system with multiple topological phases. Thus, in conclusion, edge-edge QCMI and TSE not only discriminate topological regimes from ordered phases with broken symmetries and from trivial phases, but also distinguish different types of topologically ordered phases.

\section{Systems with suppressed bulk edge correspondence: TSE of the Kitaev tie}
\label{KT}

In this section we study by means of the edge-to-edge QCMI and the quantized TSE the effects of the geometrical frustration induced by adding an hopping term of arbitrary spatial range to a Kitaev chain. The model has been originally introduced and investigated in Refs. \cite{Maiellaro2020,Maiellaro2021}.
The tight-binding Hamiltonian of the model can be written as 
\begin{equation}
H_{T} = H_K + H_d \, ,
\label{tie}
\end{equation}
where $H_K$ denotes the Kitaev chain Hamiltonian and 
\begin{equation}
H_d = - t_d(c^\dagger_{d}c_{L-d+1} + h.c.)
\label{d-hopping}
\end{equation}
is an extra long-range hopping term, connecting two symmetrical sites of the chain $d$ and $L-d+1$ with an hopping amplitude $t_d$. The long-range hopping, identified by the parameter $d$, can vary along the length of the chain, playing the role of a movable knot which induces a geometric frustration on the original chain Hamiltonian. This results in a legged-ring system with no clearly identifiable bulk, and consequently a suppressed bulk-edge correspondence, referred to as a Kitaev tie. 

The phase diagram of the system in the $\mu$--$d$ plane shows an interstitial character with topological phases nucleating inside trivial regions as the knot position $d$ is varied while keeping $\mu<2t$. This phenomenon is referred to as topological frustration.

For small values of $d$, the interstitial character of the non-trivial phases is more evident, since the system resembles a ring with very short legs. On the other hand, for large values of $d$ there is a significant growth of the topological phase domains, since the ring is reduced and the system approaches the limiting regime of a perturbed Kitaev chain. 

The Kitaev tie is realizable in single-walled carbon nanotubes \cite{Maiellaro2020,Maiellaro2021}; these are flexible ballistic conductors \cite{PhysRevLett.98.246803} where superconducting proximity effect can
be easily implemented \cite{KASUMOV1999933,PhysRevB.97.075141,PhysRevB.100.155417}.

\begin{figure}
	\includegraphics[scale=0.1]{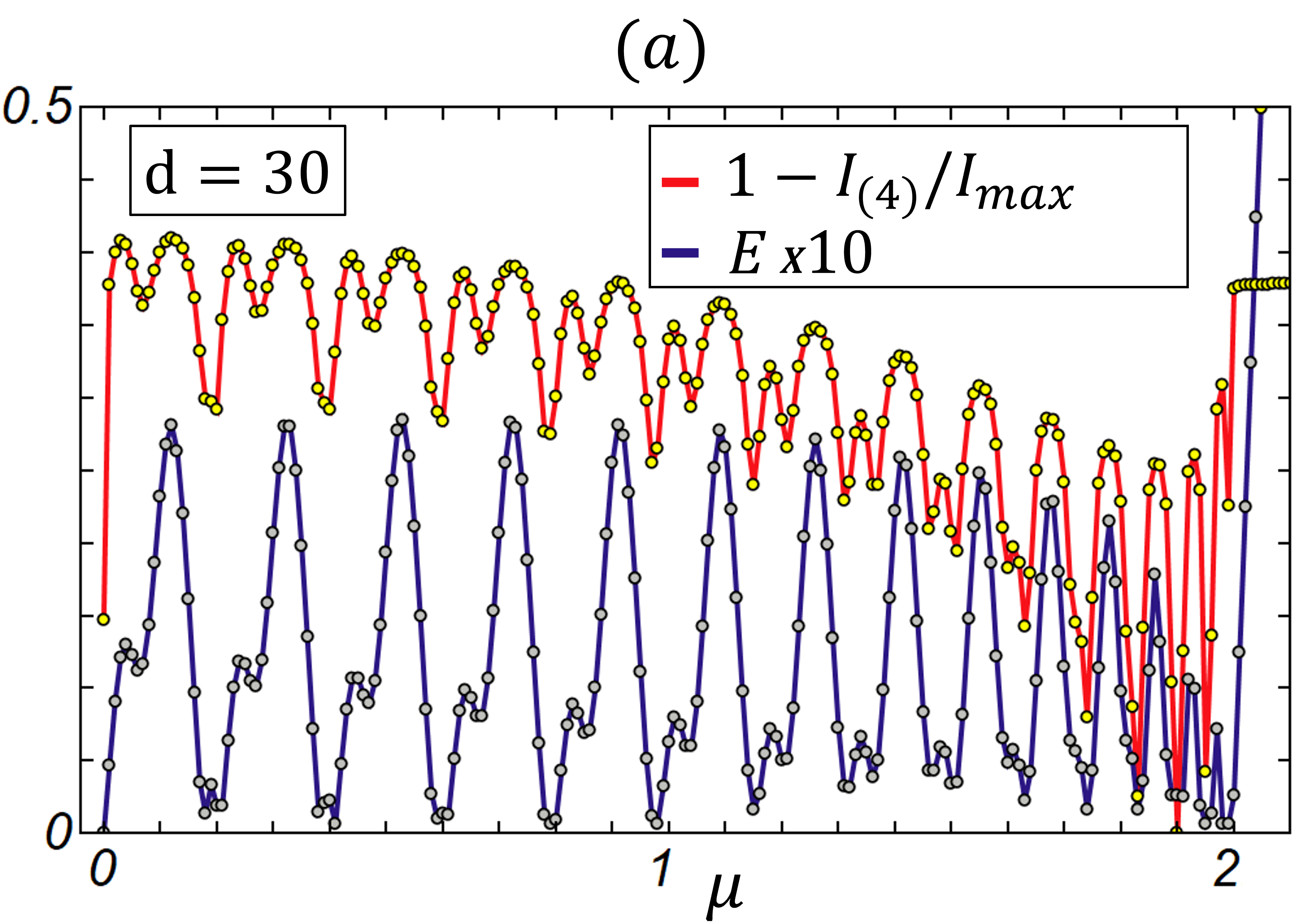}\\
	\vspace{0.3cm}
	\includegraphics[scale=0.1]{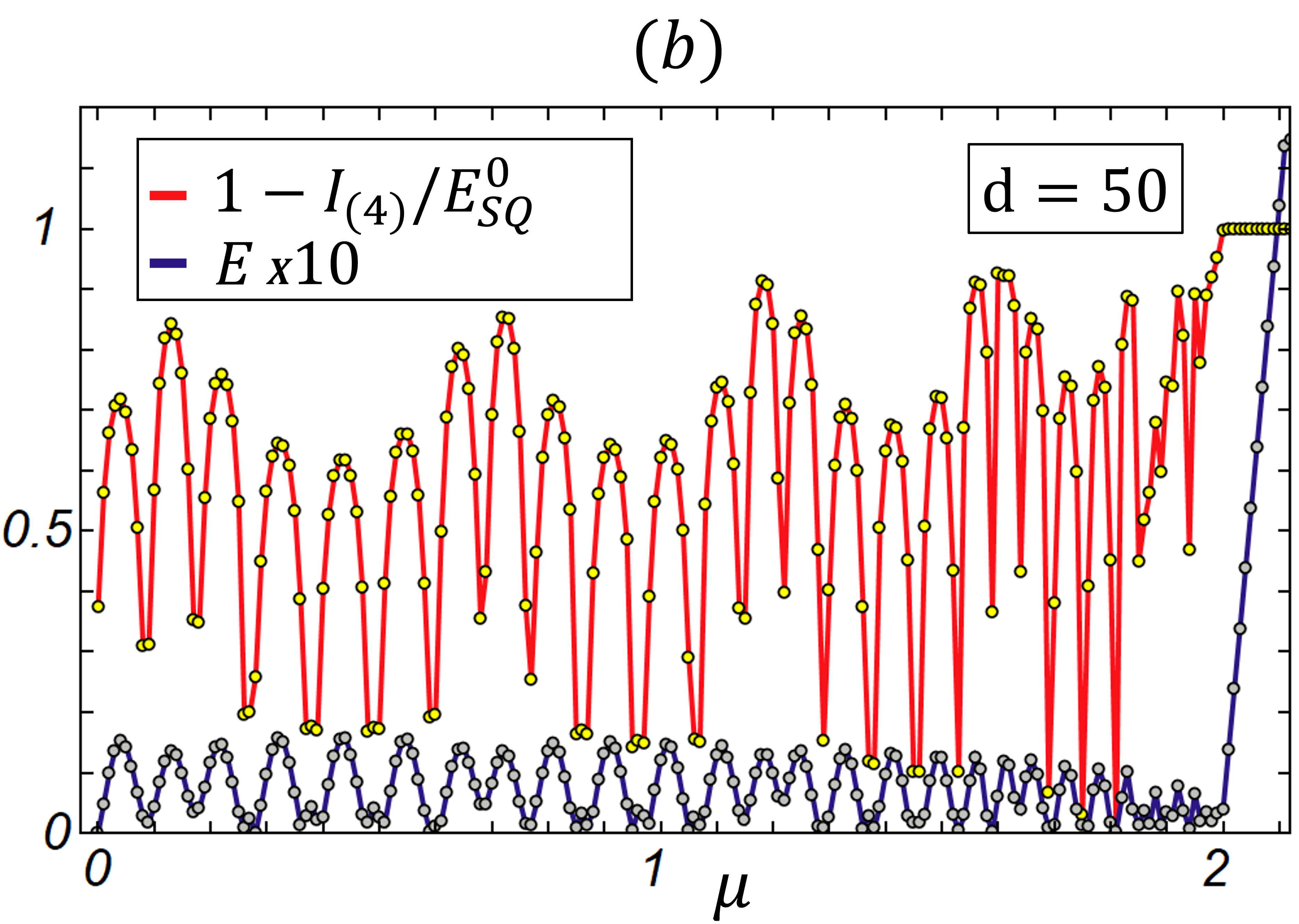}
\caption{Panel (a), red curve: behavior of the complement $ 1 - I_{(4)}/I_{max}$ of the ratio between the QCMI and its maximum value as a function of the chemical potential $\mu$. Panel (a), blue curve: behavior of the lowest energy eigenvalue $E$ as a function of $\mu$. Both curves are drawn for a Kitaev tie of $L=121$ sites, edges of $L_e=40$ sites, and the long-range hopping term positioned at $d=30$, i.e. within the system edges.
Panel (b), red curve: behavior of the complement of the QCMI-TSE ratio $1 - I_{(4)}/E^0_{sq}$ as a function of $\mu$. Panel (b), blue curve: behavior of the lowest energy eigenvalue $E$ as functions of $\mu$. Both curves are drawn for the same Kitaev tie as in panel (a), but this time with the long-range hopping positioned at $d=50$, i.e. outside the system edges. The QCMI $I_{(4)}$ is evaluated for an edge length $L_e=40$. In all plots the Hamiltonian parameters are fixed at $t=t_d$, $\Delta=0.02$ and in the unit of the hopping $t=1$.}
	\label{Figure11}
\end{figure}

In Fig.~\ref{Figure11}, we report the behavior of different indicators based on the edge-to-edge QCMI $I_{(4)}$ and the TSE $E^0_{sq}$ as functions of the chemical potential $\mu$ for a Kitaev tie of $L=121$ sites (blue curves) and we compare it with the behavior of the spectral energy function of the system (orange curves). Specifically, denoting by $I_{max}$ the maximum value achieved by $I_{(4)}$ and by $E$ the lowest energy eigenvalue, in panel (a) we compare $E$ and the complement $1 - I_{(4)}/I_{max}$ of the QCMI normalized to its maximum value as functions of $\mu$, while in panel (b) we compare $E$ and the complement of the QCMI to TSE ratio $1 - I_{(4)}/E^0_{sq}$. With respect to the extension of the edges, $L_e = L/3 = 40$, we consider two different positions of the long-range hopping. In panel (a) we fix $d=30$, i.e. with the knot included within the edges $A$ and $B$. In panel (b) we fix $d=50$, i.e. with the knot included in the traced out bulk $C_2$.

In the presence of topological frustration the occurrence of unpaired Majorana modes can be related, respectively, to minima or vanishing values of the energy eigenvalue \cite{Maiellaro2021}. Indeed, by means of Majorana polarization \cite{PhysRevB.92.115115,PhysRevLett.108.096802} it can be shown that energy minima are associated to localized Majorana modes, while energy maxima correspond to hybridized modes \cite{Maiellaro2021}. 

All plots in Fig.~\ref{Figure11} clearly show signatures of the topological frustration, pinpointed by an oscillating behavior of the various quantities as functions of the chemical potential.
For systems with an associated bulk, the QCMI $I_{(4)}$ converges to the quantized TSE $E^0_{sq}$ already for moderate lattice sizes, as shown in the discussion of the Kitaev chain and of the Kitaev ladder. 
For the Kitaev tie, the absence of an associated bulk is signaled by the fact that the complement of the QCMI to TSE ratio does not vanish, reaching instead nonvanishing minima, even for lattices of relatively large sizes ($L=121$).

On the other hand, the perfect correspondence of the positions of both the minima and the maxima of the different functions shows that the QCMI and the TSE capture the main features of the topological frustration phenomenon just as well as the energy. In particular, although the energy eigenvalue provides a stronger marker of the presence of MZEMs, remarkably the QCMI to TSE ratio $I_{(4)}/E^0_{sq}$ is a good quantifier of the emerging topological features even for systems, like the Kitaev tie, that do not allow the existence of a bulk.

Fig. \ref{Figure11} shows that the actual value of the absolute maximum $I_{max}$ reached by the QCMI $I_{(4)}$ strongly depends on whether the scope $d$ of the long-range hopping falls inside the edge extension, i.e.$d<L_e$, as in panel (a), or outside the edge extension, i.e. $d>L_e$, as in panel (b). Remarkably, $I_{(4)}$ exceeds $E^0_{sq}$ ($I_{max} \approx 0.62$) when $d<L_e$, while when $d>L_e$, the QCMI converges to the TSE $E^0_{sq}$ as in the cases of the Kitaev chains and ladders. In the case of the Kitaev tie the inapplicability of the bulk-edge correspondence is responsible for the partial loss of quantization of the QCMI that is recovered in the limit of very large scopes $d$ of the long-range hopping.

From Fig. \ref{Figure11} we see that both indicators, the energy and the QCMI to TSE ratio $I_{(4)}/E^0_{sq}$, are reliable indicators of the tie phenomenology, as both detect and capture the essentials of the topological frustration. As both indicators are sensitive to finite-size effects, the effect is stronger for the QCMI to TSE ratio: while the energy approaches the zero value, the QCMI remains within a smal but finite range from the TSE. 

\section{Discussion and outlook}
\label{Concl}

In the present work we have discussed the application of squashed entanglement to the study of one-dimensional topological systems. Squashed entanglement is a measure of bipartite entanglement defined on multipartitions that generalizes the von Neumann entanglement entropy; it allows to quantify the bipartite entanglement between any two subsystems, connected or disconnected, in any quantum state, either pure or mixed, and reduces to the von Neumann entanglement entropy on pure states. Introducing generic tripartitions and quadripartitions of a quantum system, we have identified two general classes of upper bounds on the bipartite squashed entanglement in terms of the quantum conditional mutual information between subsystems. 

For models of topological superconductors that admit Majorana zero energy modes at the system edges, we have shown that in the topological phase the upper bound provided by the edge-edge quantum conditional mutual information associated to the system quadripartition in fact coincides with the squashed entanglement between the system edges in the ground state at the exact topological point. The squashed entanglement in the topological phase is quantized at $\log(2)/2$, half the value of the maximal Bell entanglement, counting the Majorana splitting of the Dirac fermions. 

The long-distance topological squashed entanglement between the edges is constant throughout the topological phase, exhibits the correct scaling when approaching the critical point, and vanishes identically in the trivial phase. Moreover, it is stable in the presence of interactions, resilient to the effects of disorder and local perturbations, and discriminates topological superconductivity from orders associated to spontaneous symmetry breaking. It thus realizes the desired, and long-sought for, entanglement-based order parameter for symmetry-protected topological superconductivity. 

For systems with geometries of higher complexity with respect to the $1D$ Kitaev chain, featuring multiple topological phases and more than one Majorana zero energy mode per edge, like e.g. the quasi one-dimensional Kitaev ladder, topological squashed entanglement distinguishes between the multiple phases by plateaux that count the number of Majorana edge modes. For systems with suppressed bulk-edge correspondence, like e.g. the Kitaev tie, topological squashed entanglement, though ceasing to be perfectly quantized, is anyway able to identify the interstitial topological phases and discriminate them from the trivial ones. 

The long-distance TSE between the edges of a fermionic chain is reminiscent of other forms of long-distance entanglement (LDE) that are established by entanglement monogamy between the end points of one-dimensional spin-$1/2$ chains with complex interaction patterns~\cite{Giampaolo2007,Giampaolo2010,Giampaolo2015,Gualdi2011,Zippilli2013}. It is unclear at the moment whether TSE and LDE in modulated spin are related and share some common feature/origin. We plan to investigate the problem, also in connection with the intriguing possibility that complex patterns of interaction strengths might induce a kind of topological order also in some classes of spin-$1/2$ systems. Finally, we would like to remark that end-to-end LDE and Wen’s long-range entanglement patterns in the bulk are in principle quite distinct concepts; as such, making use of both of them and their interplay might lead to a deeper understanding and finer classification of topological phases of matter in higher dimensions.

The generality of the concept of squashed entanglement between arbitrary subsystems allows for several possible future research directions. Leaving aside for the moment being the tantalizing possibility of monogamy-induced multipartite extensions, here we wish to discuss the application of bipartite SE to the study of topological systems in higher dimensions as well as at finite temperature and out of equilibrium.

A crucial challenge in order to generalize the framework introduced in the present work to systems in $D \geq 2$ spatial dimensions concerns the correct identification of the different bulk and edge parts in the multipartitions. Here we wish to mention that among topological two-dimensional systems, second order topological superconductors ($HOTSC_2$) \cite{Maiellaro2021bis,Maiellaro2022,PhysRevLett.122.236401,PhysRevResearch.2.033495} are of particular relevance both for their fundamental properties and for the possibility that they offer to implement braiding dynamics in a rather straightforward way \cite{PhysRevResearch.2.032068}. Second-order topological superconductors are two-dimensional systems with gapped one-dimensional boundaries and zero-dimensional localized modes (corner modes). For these systems, the identifications of edges $A$, $B$ and bulks $C$ $C_1$, $C_2$ is clear and easily connected to that of the one-dimensional case. This class of systems represents provides a suitable arena to generalize TSE, albeit numerically, in $D = 2$ dimensions. We plan to report on these and related problems in upcoming future work.

In order to extend even further the concept of edge QCMI and edge SE to the full generality of arbitrary systems in any spatial dimension, the central issue concerns the correct identification of the system edges, their localization properties, and their degree of connectivity or lack thereof. Clearly, this type of analysis will be heavily model-dependent. For instance, regarding the role of open boundary conditions in higher dimensions, identification of the bulk and of the edges will depend on the form of the boundaries, for instance on the number and form of the corners and vertices for different lattice geometries.

When considering periodic or partially periodic boundary conditions, including e.g. ring, cylinder, and torus geometries, the straightforward choices appear to be tripartitions and quadripartitions of the system in adjacent arches in order to realize and isolate effectively disconnected edges. To set the stage, consider for instance a ring geometry fixed by imposing periodic boundary conditions (PBCs). In such a geometrical configuration, the two edge Majorana Zero Energy Modes (MZEMs) of the original open chain recombine with each other and annihilate in a full fermion. Despite the fact that MZEMs are no longer present, the topological invariants continue to yield signatures of the topological nature of the system, provided one suitably modifies the choice of partitions. Considering an elongated ring (think for instance of an autodrome), in order to realize a tripartition suitable for long-distance entanglement (LDE) of the topological squashed entanglement (TSE) type, we need to cut the two antipodal curves that will then play the role of the long-distance subsystems $A$, $B$, and consider as the bulk $C$ the two remaining disconnected segments connecting the two arches $A$, $B$. By taking each of the two segments separately, i.e. as parts $C_1$, $C_2$ of the full bulk $C$, we can then realize the desired quadripartition as well.

Finally, in a very trivial sense, given that on pure quantum states, e.g. ground states of many--body systems, the squashed entanglement reduces exactly to the von Neumann block entanglement entropy when the system is partitioned into just two simple blocks, the subleading contribution to the von Neumann block entanglement entropy, the block topological squashed entanglement, trivially coincides with the topological entanglement entropy in the case of true topological order in two-dimensional systems. Restricted to this obvious meaning, squashed entanglement already provides a framework unifying different classes of topological order in higher dimensions.

Another relevant line of research concerns the generalization of TSE to open topological systems, i.e. topological systems coupled to an environment. In this case a first possible step forward can be based on the self-energy formalism, by reducing the original model to an effective system structurally equivalent to the original but where the edges and the bulk are renormalized by the effects of the environment. Indeed, in such case it is possible to recover the full partition structure of a closed system. We plan to apply this method, via the corresponding non-Hermitian Hamiltonians, to the investigation of edge QCMI and edge SE for open topological systems in one dimension.

Extending further the above line of research leads to the problem of applying QCMI and SE to the study in full generality of topological quantum matter at finite temperature and out of equilibrium \cite{Cooper2019}. In fact, even for non topological matter it would be interesting to study finite temperature quantum criticality resorting to squashed entanglement and compare it systematically with other measures of nonclassicality such as, e.g., the entropic discord \cite{Illuminati2021} or the Bell nonlocality \cite{Illuminati2022}. In this perspective, hierarchies of quantum complexity according to different layers of quantumness, ranging from discord and coherence to entanglement and nonlocality, could realize the scaffold of a unified research framework applicable to a wide spectrum of problems, ranging from quantum information and quantum matter to elementary particle physics and quantum gravity. 

{\it Acknowledgments} -- We acknowledge support by MUR (Ministero dell’Università e della Ricerca) via the project PRIN 2017 ”Taming complexity via QUantum Strategies: a Hybrid Integrated Photonic approach” (QUSHIP) Id. 2017SRNBRK. 
We thank an anonymous referee for several enlightening comments that helped to improve on the original manuscript.

{\it Author contributions} -- F.I. conceived the fundamental idea, organized the project and supervised the work. F. I. and A. Marino performed part of the calculations, while A. Maiellaro performed the full calculations, produced the numerical plots and realized the figures. F. I. and A. Maiellaro drafted the manuscript, discussed the results and edited the paper.

\bibliography{Bibtex_new}

\end{document}